\documentstyle[preprint,aps,prb,epsf,tighten,eqsecnum]{revtex}
\def\epsfsize#1#2{\epsfxsize}
\begin{document}
\draft
\title{The superconductor-insulator \\
transition in 2D dirty boson systems}
\author{Mats Wallin}
\address{Department~of~Theoretical~Physics,
Royal~Institute~of~Technology, S-100~44~Stockholm, Sweden}
\author{Erik~S.~S\o rensen}
\address{Department~of~Physics,
University of British Columbia,
Vancouver, BC, V6T 1Z1, Canada}
\author{S.~M.~Girvin}
\address{Department~of~Physics,
Indiana~University, Bloomington, IN~47405}
\author{A.~P.~Young}
\address{Department~of~Physics,
University of California, Santa Cruz, CA~95064}
\date{September 20, 1993}
\maketitle

\begin{abstract}
Universal properties of the zero temperature superconductor-insulator
transition in two-dimensional amorphous films are studied by extensive
Monte Carlo simulations of bosons in a disordered medium.
We report results for both short-range and
long-range Coulomb interactions for several different points in
parameter space. 
In all cases we observe a transition from a
superconducting phase to an insulating Bose glass phase. 
From finite-size scaling of our Monte Carlo data we determine 
the universal conductivity $\sigma^*$ 
and the critical exponents at the transition.
The result $\sigma^* = (0.55 \pm 0.06) (2e)^2/h$ for bosons with 
long-range Coulomb interaction is roughly consistent with experiments
reported so far.
We also find $\sigma^* = (0.14 \pm 0.03) (2e)^2/h$ for
bosons with short-range interactions.
\end{abstract}

\section{Introduction}

From the work of Abrahams, Anderson, Licciardello, and
Ramakrishnan~\cite{gangof4} it is known that no true metallic behavior
can be observed for non-interacting electrons at $T=0$ in two
dimensions, since all states
will be localized by arbitrarily weak disorder. 
When repulsive interactions are turned on the situation is less clear but the
general belief \cite{lr85}
is that a metallic phase still should be absent at $T=0$
in the presence of disorder, although we know of no rigorous proof
of this.
However, in the presence of attractive interactions,
a superconducting phase is expected~\cite{comment0}, both at $T=0$ and
finite $T$, even for a finite amount 
of disorder, because disorder is irrelevant~\cite{comment-1}
at the finite temperature
transition, which is of the Kosterlitz-Thouless~\cite{KT}
type discussed below.
The onset of superconductivity at $T = 0$
is presumed then, in $d = 2$, to be directly from
the insulating phase with no intervening metallic phase.
One should therefore in principle be able to observe a direct
insulator-superconductor transition at zero temperature in
two dimensions as a function of disorder and/or interaction strength. 
The main topic of this paper is to analyze such a transition
and extract its universal features.

Dimensionality and divergent length scales play an important 
role in continuous phase transitions.
The diverging correlation length scale implies
that many microscopic details are irrelevant.
Furthermore physical quantities containing dimensions of
length to some non-zero power typically diverge or 
vanish at the critical point.
Two dimensions is special in that
the conductivity contains no length scale units, i.e.\ the
conductance per square is the same as the conductivity.
Hence, right
at the $T=0$ quantum critical point, the conductivity
is not only finite and nonzero but also
{\em universal}\,\cite{fisher90a,wen},
even though it is zero in the insulating
phase and infinite in the superconducting phase.
This view differs from that of
previous work~\cite{chakravarty} which parameterized the transition in
terms of the {\it normal} state resistivity.
The calculation of this universal conductivity is one of the main
goals of the present paper.
A short account on some of our results has already been
published~\cite{sorensen92}.

\begin{figure}[htb]
\centering
\epsfysize=14 cm
\leavevmode
\epsffile{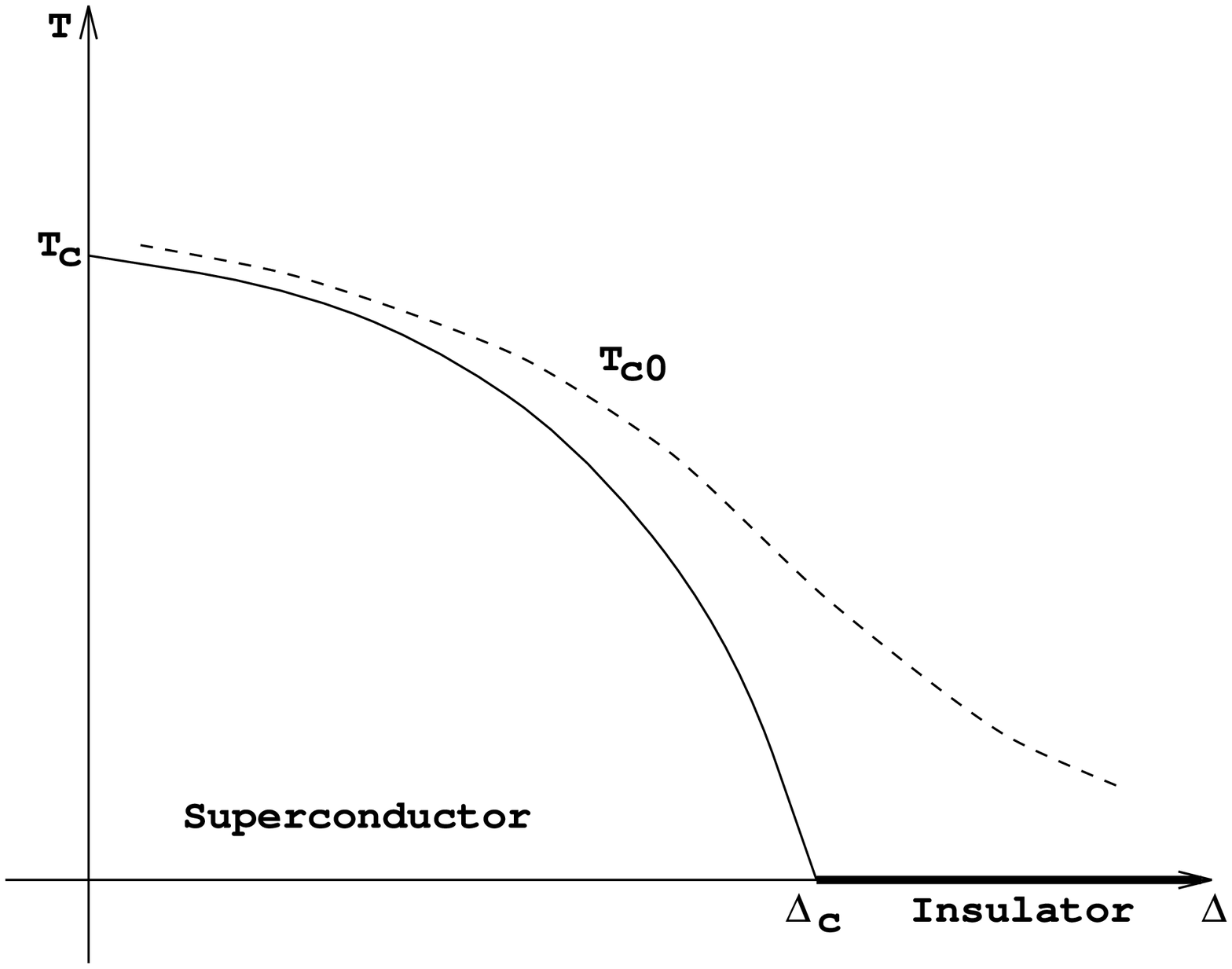}
\caption{
Sketch of the general phase diagram for thin-film superconductors
as a function of disorder, $\Delta$, and temperature, $T$.
The dashed line, $T_{c0}$,
indicates the mean-field onset temperature where Cooper pairs
start to form. At $T=0$ an insulating phase
appears (solid line) and a transition from a superconductor to
an insulator takes place at a critical value of the disorder
$\Delta_c$.}
\label{fig:scphase}
\end{figure}

A schematic phase diagram is shown in Fig.~\ref{fig:scphase} as
a function of temperature, $T$, and disorder, $\Delta$.
At zero temperature, a critical amount of disorder, $\Delta_c$,
separates the superconducting from the insulating phase. 
Even at finite temperatures
the superconducting phase persists whereas a truly insulating
phase only exists at $T=0$, because, at finite $T$, electrons can be
{\em inelastically} scattered from one localized state to another, and
hence conduct.
This insulating phase, consisting of localized electron pairs,
can then be described, close to the critical point, as a 
Bose-condensed fluid of vortices. The universality class of the
transition should therefore be that of the superconductor to
Bose glass~\cite{fisher89c}.

Let us first discuss the nature of the transition at finite temperatures
indicated by the solid line in Fig.~\ref{fig:scphase}. 
The finite temperature transition should have many similarities
with the 2D XY transition at which logarithmically interacting
vortices unbind~\cite{KT}. However, Pearl~\cite{pearl} showed
that in a superconducting film vortex pairs only have logarithmic
interactions out to a distance $\Lambda_\perp=2\lambda^2/d$ beyond which
the interaction energy falls off as $1/r$. Here $\lambda$ is the bulk
penetration depth, $d$ the film thickness, and $\Lambda_\perp$ the
screening length for magnetic fields. Due to this
cutoff, the energy required to create a vortex is always {\it finite}
and no sharp transition should exist. However, according to the
Kosterlitz-Thouless theory~\cite{KT}, the value of $\Lambda_\perp(T_c)$
is given {\em exactly} by  \cite{halnelson,DH,BMO}
$ \Lambda_\perp(T_c) =  \phi_0^2 / (16 \pi^2 k_B T_c)$,
where $\phi_0 = hc / 2e$ is the flux quantum. Numerically
$\Lambda_\perp(T_c) = 2 / T_c$ where $T_c$ is in Kelvin and 
$\Lambda_\perp(T_c)$ is in {\em centimeters}. Thus, $\Lambda_\perp$
is so large at $T_c$ that
rounding of the transition due to the presence of free vortices below
$T_c$ is almost certainly unobservable. In fact, rounding due to
finite-size effects is probably more important.

The vortex unbinding transition at $T_c$ is
driven by fluctuations in the phase of the superconducting
order parameter.
At a higher temperature, $T_{c0}$, fluctuations in the amplitude
of the order parameter will become important and a crossover
to a regime dominated by paraconductivity~\cite{skocpol} will occur.
$T_{c0}$ is indicated by the dashed line in Fig.~\ref{fig:scphase}.
For $T_c\le T \ll T_{c0}$ the presence of free vortices destroys
the characteristic global properties of the superconducting phase.
Nevertheless, 
a {\it local} order parameter still exists between $T_c$ and $T_{c0}$. 
The presence of free vortices leads to a finite conductivity
of the form~\cite{halnelson} $\sigma_v\sim 0.37\sigma_n(\xi/\xi_c)^2$,
where $\sigma_n$ is the conductivity of the normal state electrons,
$\xi_c$ the core size of a vortex and $\xi$, a typical distance between
free vortices,
is the Kosterlitz-Thouless~\cite{KT}
correlation length which diverges exponentially at
$T_c$. The exponential tail in the resistivity caused by the presence
of free vortices between $T_c$ and $T_{c0}$ has been observed
experimentally~\cite{fiory,kadin} in superconducting films with
high normal state resistivity. In these experiments the ``mean field
onset temperature'', $T_{c0}$ is determined by fitting the resistance
to an Aslamazov-Larkin~\cite{AL} form, and $T_{c0}-T_c$ is found to
be of the order of half a Kelvin. In the dirty limit,
Beasley et al.~\cite{BMO} derive the
relationship $\tau_c\equiv (T_{c0}-T_c)/T_c\sim 0.17 e^2/\sigma_n\hbar$
so, for films with a relatively high sheet resistance, $\tau_c$ can 
be appreciable.  For a review of the finite temperature transition we
refer the reader to Refs.~\onlinecite{mooij}.
More recently some evidence for a vortex-antivortex unbinding transition
in superconducting niobium films~\cite{hsu} with $R_\Box=122\Omega$
has been found. However,
the difference
between $T_c$ and $T_{c0}$ is very small in the clean limit
which makes the
Kosterlitz-Thouless behavior difficult to observe.

The superconducting order parameter is a complex scalar,
described by both a magnitude and a phase. 
Our key basic assumption 
is that universal properties at superconductor-insulator transition are
determined
only by phase fluctuations, as outlined above, and that the magnitude
of the order parameter, and therefore of
the gap in the fermionic energy spectrum, 
remains finite at the critical point.
We thus assume that when disorder drives $T_c$ to zero,
$T_{c0}$ remains nonzero.
If 
the transition is approached from the insulating side, 
a local order parameter appears
before the onset of global phase coherence at $\Delta_c$.
Implicit in our assumption is that Cooper pairs,
and thus a gap to fermionic excitations, persists into the insulating
phase, even though superconductivity is destroyed by phase fluctuations.
On the scale of a diverging 
{\em phase}-correlation length, $\xi$, the
individual Cooper pairs will look like point particles. The fermionic
degrees of freedom should therefore be highly suppressed at the critical
point and an approximate description in terms of point-like bosons
should be valid. 
It is possible that strong disorder
destroys the local fermionic gap at a finite density of points, but,
provided that the Fermi degrees of freedom are localized,
they may still be dynamically irrelevant and
our model applicable.

Since we shall be concerned with the transition
at $\Delta = \Delta_c, T=0$,
vortices in the system will not be excited thermally, 
but there will be vortices present created by 
quantum fluctuations~\cite{RANA-GIRVIN}.
We therefore need to treat the vortices as quantum mechanical 
objects and one might expect the transition at $\Delta_c$ to be
described by a (2+1)D XY model~\cite{MHL}, where the
extra dimension arises because we are considering a
T=0 quantum phase transition, see e.g.\ Ref.~\onlinecite{fisher89c}. 
However as we shall see in Section~\ref{sec:boshub}, although the physics is
indeed described by a (2+1)D system, its symmetry is, in general,
{\em not} that of the XY model.
It is also worth noting that the vortex mobility, 
$(2e^2/\pi\hbar^2)\xi^2R_\Box$~\cite{mooij}, is significantly augmented
in dirty superconducting films. Hence, at $\Delta_c$ the vortices should
be seen as fairly light objects that move rather freely. At still higher
disorder the Bose glass phase should cross over into an Fermi
glass when the individual electrons constituting the bosons become
localized. This behavior 
may have already been  observed in a magnetic field~\cite{paalanen92}.

Recent experiments seem to confirm that a direct
insulator-superconductor transition indeed does take place at zero
temperature in many materials.
Haviland et al.~\cite{haviland} and Liu et al.~\cite{liu} have 
performed experiments on Bi films grown in situ. 
The experimental technique is described in Ref.~\onlinecite{jaeger}. 
These films are believed to be truly amorphous on an atomic scale. 
The authors report a 
critical d.c.\ resistivity, $R_\Box^\ast$,
very close to $R_Q$, where 
\begin{equation}
R_Q=h/4e^2 \approx 6453 \Omega \ .
\label{rq}
\end{equation}
Furthermore, experiments performed on DyBaCuO films~\cite{wang91,wang92}
and NdCeCuO~\cite{tanda} show clear evidence for a direct 
superconductor-insulator transition.
The reported critical resistivity seems in this case to be somewhat 
higher, around 10 k$\Omega$ or 1.5 $R_Q$~\cite{wang91,wang92}.
Lee and Ketterson~\cite{lee90} have presented results from
experiments on MoC films again showing very clear evidence for
a superconductor-insulator transition occurring at zero temperature,
but with $R_\Box^\ast$ slightly lower, in the range 2.8-3.5 k$\Omega\ \sim
0.5 R_Q$. Furthermore, experiments performed on Josephson junction
arrays~\cite{geerligs,zant}, which are believed to be in the
same universality class, also seem to support the picture
of a superconductor-insulator transition. The existence of a
superconductor-insulator transition in two-dimensional films at zero
temperature thus seems well established
but 
evidence for the universality of the critical resistivity
remains weak. It is 
not clear, however, whether all the experiments are in the critical region.
In order to establish that a given experiment is actually probing the critical 
regime, one must show scaling of
the resistivity data.
This has been done successfully by Hebard and
Paalanen~\cite{hebard90,hebard92}
for the field-tuned transition and 
partially successfully by the Minnesota group~\cite{GOLDMANscaling}.
However it is likely that most measurements to date have failed to 
probe the critical regime,
and further experiments at even lower temperatures 
are expected to give better
agreement among the different estimates of $R_\Box^\ast$.

The situation concerning the relevance of a bosonic picture
seems less clear. Hebard and Paalanen~\cite{hebard90,hebard92}
have reported results on amorphous InO films in a magnetic field, 
supporting the existence of Cooper pairs
in the insulating phase. 
For the $B=0$ transition, Hebard and Paalanen~\cite{HP85}
have presented clear transport evidence that $T_{c0}$ 
remains finite as $T_c$
is driven to zero.
On the other hand, direct tunneling
measurements by Dynes et al.~\cite{dynes} and Valles et al.~\cite{valles}
on homogeneously disordered Pb films shows
that the gap goes to zero at the critical point. There seems, however,
to be a general agreement that a local superconducting order parameter
exists prior to the transition in granular films and in Josephson junction
arrays where the individual grains become superconducting above $T_c$.
It is possible that tunneling experiments tend to emphasize 
regions of the samples containing quasilocalized fermion 
states below the gap which are necessary
to achieve tunneling.

A number of theoretical and numerical studies of the 
superconductor-insulator transition have been performed. 
Gold~\cite{gold} studied
the impurity induced insulating transition
in the interacting Bose gas. Giamarchi and Schulz~\cite{giamarchi87}
considered the one-dimensional electron gas with attractive interactions
in the presence
of disorder. They found a transition to a localized phase in the same
universality class as that of repulsively interacting bosons in a random
potential.  
This lends strong support to our
assumption of the dirty boson universality class in the 2D case.
Fisher et al.~\cite{fisher88b,fisher88a,fisher89a,fisher89c} considered
the boson Hubbard model and, through a scaling analysis,
derived equations for the exponents governing the 
superconductor-insulator transition 
as well as the phase diagram for dirty bosons.
A renormalization group approach was taken by 
Weichman et al.~\cite{weichman88}
who performed a double epsilon expansion for the dirty boson problem.
Following the initial suggestion of a Bose glass phase
in the disordered system and a Mott insulator in the clean system,
Batrouni et al.~\cite{batrouni90} and Krauth et al.~\cite{krauth91a},
showed,
by quantum Monte Carlo simulations, 
the existence of Mott insulating phases in an interacting boson
system without disorder, characterized by the exponents predicted by 
Fisher et al.~\cite{fisher89c}
Subsequently, these authors considered the disordered case
and evidence for a Bose glass was found.\cite{scalettar91,krauth91b}
A Bose glass phase was also observed in a real space
renormalization group study by Singh et al.~\cite{singh}
The universal conductivity was first calculated by a $1/N$ expansion 
and Monte Carlo methods for the (2+1)D XY model by Cha 
et al.~\cite{cha91} and Girvin et al.~\cite{nobel,progtheo}
The universal conductivity for disordered
bosons was then calculated by Runge~\cite{runge} by exact
diagonalization techniques on small lattices. 
Universal properties 
for a boson system in the presence of disorder both with and without
long-range interactions were calculated by S\o rensen 
et al.~\cite{sorensen92,thesis} by Monte Carlo simulations, using a path
integral representation which, effectively, only includes phase
fluctuations in the Bose field.
A universal conductivity was also recently found by Kampf 
et al.~\cite{kampf} in the boson Hubbard model including both phase and
amplitude fluctuations. 
Two recent works have recently been
published after the present work was finished. 
Batrouni et al.~\cite{batrouni93}
have calculated the universal conductivity by quantum Monte Carlo
simulations directly on the boson Hubbard model, and
Makivi\'c et al.~\cite{makivic93} have calculated the exponents
and the universal conductivity using a hard-core boson model. The
results of these last two papers differ from ours, and we shall comment
on this in section \ref{discuss}.

Here we shall consider two forms of interaction between the bosons:
short-range repulsive interaction and 
long-range Coulomb interaction.
The model with short-range interactions
is relevant to experiments on the onset of superfluidity in
$^4$He films~\cite{cha91}.
However, our present results for the zero temperature transition in 2D are not
directly applicable to He experiments
in porous media such as Vycor or xerogel~\cite{chan,wong}, since
these experiments are mainly concerned with the 3D transition
at finite temperatures. As stated above, the
model with Coulomb interactions is expected
to be in the correct universality class to describe the 
superconductor-insulator transition.
However, the model
and many of the results presented in this paper
are applicable to other systems too.
The world lines of the dirty boson model describe a gas of stringlike
objects in a random medium.
In addition to the superconductor-insulator transition
this model may also apply to other problems such as
vortex lines in high-temperature superconductors
with correlated pinning centers~\cite{nelson-vinokur,MW-SMG}, 
and polymer solutions.
Our results for {\em universal} quantities might also be relevant for 
these problems.

The organization of the paper is as follows. 
In Section~\ref{sec:boshub} we shall construct a form
of the boson Hubbard model, including disorder and interactions,
which is suitable for Monte Carlo simulations.
Here we shall assume, as discussed above,
that only bosonic degrees of freedom, 
i.e.\ complex order parameter fluctuations,
are relevant at the superconductor-insulator transition. 
In addition, to further
simplify the numerical work, we effectively include
only phase fluctuations of the bosons, amplitude fluctuations being
neglected.
Section~\ref{sec:scaling} describes the scaling theory of the quantities
that we are interested in. We discuss, in Section~\ref{sec:fss},
how we determine these quantities in the simulation, and we also treat
the finite-size scaling techniques
needed to extrapolate our results to infinite size.
Section~\ref{sec:methods} describes our Monte Carlo methods, while
Section~\ref{sec:dirtyb} presents our results for short-range interactions
and disorder, relevant to experiments on helium films. 
In Section~\ref{sec:lr} long-range Coulomb interactions 
are included along with disorder.
We believe that
this model contains all ingredients necessary to make it
relevant to experiments on the superconductor-insulator transition; i.e.,
that it is in the correct universality class. Our results are discussed
in Section \ref{discuss}.

\section{The model}\label{sec:boshub}

In this section we introduce our basic model and via a sequence of
transformations arrive at a form suitable for
Monte Carlo simulation.
As argued above it should be possible to describe the
universal features of the superconductor-insulator transition 
in terms of boson physics. In this section we
shall argue that the relevant starting point is the boson Hubbard
model with a random local chemical potential (site energy).
If only phase fluctuations are
relevant we can map this model onto a dual Villain type model. 
We shall see that 
only in the absence of disorder and when there is
an integer number of bosons per site, does
this model belong to the same universality class as
the (2+1)D XY model.

In order to model the superconductor-insulator transition in terms
of bosons, we must include an on-site repulsive interaction,
otherwise all bosons would collapse into the lowest lying, 
highly localized state.
The on-site repulsion term is the simplest possible
way to model Coulomb repulsion.  
The correct treatment of the long-range part
of the interaction will be discussed below.
For simplicity we shall take an underlying square lattice of 
spatial size $N = L \times L$.
Changes in the symmetry of the lattice are not expected to modify the
critical behavior of the model.
We can then write down the boson Hubbard model in presence of 
disorder\cite{fisher89c,fisher89a,fisher89b}:
\begin{equation}
H_{\text{bH}} = H_0+H_1
\end{equation}
where
\begin{eqnarray}
H_0&=&\frac{U}{2}\sum_{\bf r}\hat n_{\bf r}^2 -
\sum_{\bf r}(\mu+v_{\bf r} - zt)\hat n_{\bf r}\nonumber\\
H_1&=& - t\sum_{\langle {\bf r},{\bf r'}\rangle }
(\hat \Phi^\dagger_{\bf r}\hat \Phi_{\bf r'}+
\hat \Phi_{\bf r}\hat \Phi^\dagger_{\bf r'}) \ .
\label{eq:bosonhubbard}
\end{eqnarray}
Here $U$ is the on-site repulsion,
$\mu$ is the chemical potential, z the number of nearest neighbors,
and $v_{\bf r}$ represents the random on-site potential varying
uniformly in space between $-\Delta$ and $\Delta$. 
As usual, $\hat n_{\bf r}=\hat \Phi^\dagger_{\bf r}\hat \Phi_{\bf r}$
is the number operator on site ${\bf r}$.
The hopping strength is given by $t$, 
and $\langle {\bf r},{\bf r'}\rangle $
denotes summation over pairs of nearest neighbors,
each pair counted once.

In the absence of disorder there is no insulating phase unless we 
fix the boson density at an 
integer value, $n_0$. Let us consider this case first.
If we set $\hat \Phi_{\bf r}\equiv|\hat \Phi_{\bf r}|e^{i\hat \theta_{\bf r}}$
and integrate out amplitude fluctuations,
the boson Hubbard model, Eq.~(\ref{eq:bosonhubbard}),
becomes a model of coupled Josephson junctions,~\cite{fisher89c,fisher88b}
\begin{equation}
H_{\text{JJ}}=\frac{U}{2}\sum_i\hat n_{\bf r}^2
-\sum_{\langle {\bf r},{\bf r'}\rangle }
t \cos(\hat\theta_{\bf r}-\hat\theta_{\bf r'}) \ ,
\label{eq:boshub}
\end{equation}
where, in this representation,
$\hat n_{\bf r}$, which denotes the deviation of the boson number from $n_0$,
runs from $-\infty$ to $\infty$ and so Eq.~(\ref{eq:boshub})
can only be quantitatively compared with the Hubbard model,
Eq.~(\ref{eq:bosonhubbard}),
when $n_0$ is very large, but is expected to be in the same universality
class for arbitrary integer $n_0$. Note that $t$ in Eq.
(\ref{eq:boshub}) is $2 n_0 $ times the parameter $t$ in Eq.
(\ref{eq:bosonhubbard}). The phase operator, $\hat\theta_{\bf r}$ is
canonically conjugate to $\hat n_{\bf r}$ so
this version of the boson Hubbard model can be written in the angle
representation as
the quantum rotor model~\cite{cha91,fisher88b,fisher89c,doniach}
\begin{equation}
H_{\text{qr}}={U \over 2}\sum_{\bf r} 
\left( \frac{1}{i}\frac{\partial}{\partial
\theta_{\bf r}} \right)^2
-\sum_{\langle {\bf r},{\bf r'}\rangle }
t \cos(\theta_{\bf r}-\theta_{\bf r'}) \ .
\label{hqr}
\end{equation}

Let us write the partition function corresponding to $H_{\text{qr}}$ as
\begin{equation}
Z = {\rm Tr}\, \exp[-\beta(T+V)] \ , 
\label{eq:three}
\end{equation}
where the kinetic energy of the rotors is
\begin{equation}
T = -{U \over 2}\sum_{\bf r}\frac{\partial^2}{\partial\theta_{\bf r}^2} \ ,    
\end{equation}
(which corresponds to the potential energy of the bosons) and the
potential energy of the rotors is 
\begin{equation}
V = -\sum_{\langle {\bf r},{\bf r'}\rangle }
t \cos(\theta_{\bf r}-\theta_{\bf r'}) \ .
\end{equation}

We evaluate the trace in the partition function by 
writing a path integral over
$M$ time slices $\tau_j$ between $\tau = 0$ and $\tau = \beta$:  
\begin{eqnarray}
Z &=& {\rm Tr}\left\{\exp[-\beta(T+V)]/M\right\}^M\nonumber\\
& = & \lim_{M\to\infty}
{\rm Tr}\left\{\exp[-\Delta\tau\ T]\exp[-\Delta\tau V]\right\}^M \ ,
\label{eq:seven}
\end{eqnarray}
where $\hbar \tau$ is imaginary time and
\begin{equation}
\Delta\tau = \beta / M
\end{equation}
is the width of one time slice. Note that the limit $\Delta\tau \to 0$
must be taken to correctly represent the underlying quantum mechanics
problem.
Eq.~(\ref{eq:seven}) can be rewritten by inserting complete sets of states
\begin{equation}
Z \approx \int{\cal D}\theta \prod_{j=0}^{M-1}
\langle\{\theta(\tau_{j+1})\}| \exp[-\Delta\tau T]\exp[-\Delta\tau V]
|\{\theta(\tau_j)\}\rangle \ ,
\label{eq:eight}
\end{equation}
where $|\{\theta(\tau_j)\}\rangle$ is a coherent state 
in which site ${\bf r}$
has phase $\theta_{\bf r}(\tau_j)$ at time $\tau_j$ 
and the trace is enforced by periodic boundary conditions
\begin{equation}
\{\theta(\tau_M)\} = \{\theta(\tau_0)\} \ .
\label{eq:nine}
\end{equation}
The coherent states are eigenstates of the potential so
\begin{equation}
\exp[-\Delta\tau V] \Big|\{\theta(\tau_j)\}\Big\rangle =
\exp\left\{\Delta\tau t \sum_{\langle 
{\bf r,r^{\prime}}\rangle}
\cos\left[\theta_{\bf r^{\prime}}(\tau_j) 
- \theta_{\bf r}(\tau_j)\right]\right\}
\Big|\{\theta(\tau_j)\}\Big\rangle \ ,
\label{eq:10}
\end{equation}
where the sum is over all nearest neighbor spatial pairs,
and hence Eq.~(\ref{eq:eight}) becomes
\begin{equation}
Z\approx \int{\cal D}\theta\prod_{j=0}^{M-1}
\exp\left\{K_x \sum_{\langle 
{\bf r,r^{\prime}}\rangle}
\cos[\theta_{\bf r^{\prime}}(\tau_j) 
- \theta_{\bf r}(\tau_j)]\right\} T_j \ ,
\label{eq:11}
\end{equation}
where
\begin{equation}
T_j\equiv\langle\{\theta(\tau_{j+1})\}
|e^{-\Delta\tau T}|\{\theta(\tau_j)\}\rangle \ ,
\label{eq:12}
\end{equation}
and
\begin{equation}
K_x = t \Delta\tau \ .
\label{Kx}
\end{equation}
Since the kinetic energies on different sites commute, 
we can consider each site separately:
\begin{equation}
T_j = \prod_{\bf r}\left\langle
\theta_{\bf r}(\tau_{j+1})\left|\exp\left[
\frac{\Delta\tau U}{2}\frac{\partial^2}{\partial\theta^2_{\bf r}}\right]
\right|\theta_{\bf r}(\tau_j)\right\rangle \ .
\label{eq:13}
\end{equation}
Let $J^\tau_{\bf r}(\tau_j)$ be the integer-valued angular momentum 
at ${\bf r}$ at time $\tau_j$.  
The corresponding state has wave function
\begin{equation}
\langle\theta_{\bf r}(\tau_j)|J^\tau_{\bf r}(\tau_j)\rangle =
e^{iJ^\tau_{\bf r}(\tau_j)\theta_{\bf r}(\tau_j)} \ ,
\label{eq:14}
\end{equation}
which is an eigenfunction of the kinetic energy.  Inserting this complete set
of states, we have
\begin{equation}
T_j = \sum_{\{J\}}\prod_{\bf r}\Big\langle
\theta_{\bf r}(\tau_{j+1})\big|J^\tau_{\bf r}(\tau_j)
\Big\rangle
\exp\left\{-\frac{\Delta\tau U}{2}\left[J^\tau_{\bf r}(\tau_j)\right]^2\right\}
\Big\langle J^\tau_{\bf r}(\tau_j)\big|\theta_{\bf r}(\tau_j)\Big\rangle \ ,
\label{eq:15}
\end{equation}
and thus
\begin{eqnarray}
Z\approx \int{\cal D}\theta \sum_{\{J\}}
& \exp & \left\{K_x \sum_{\langle 
{\bf r,r^{\prime}}\rangle}\sum_{j=0}^{M-1}
\cos\left[\theta_{\bf r^{\prime}}(\tau_j) 
- \theta_{\bf r}(\tau_j)\right]\right\}\nonumber\\
\times & \exp & \left\{-\frac{\Delta\tau U}{2}
\sum_{\bf r}\sum_{j=0}^{M-1}
\left[J^\tau_{\bf r}(\tau_j)\right]^2\right\}\nonumber\\
\times & \exp & \left\{i\sum_{\bf r}\sum_{j=0}^{M-1}
J^\tau_{\bf r}(\tau_j)\left[\theta_{\bf r}(\tau_j) 
- \theta_{\bf r}(\tau_{j+1})\right]\right\} \ .
\label{eq:16}
\end{eqnarray}
We can now proceed in two possible ways.  
We can either integrate out the angular variables $\{\theta\}$ to obtain 
a statistical mechanics problem in the integer variables $\{J\}$, 
or we can sum over the $\{J\}$ to obtain a classical
(2+1)-dimensional XY model.  Let us start with the latter.

Because $\Delta\tau$ is small, the sum over the 
$\{J\}$ is slowly convergent.  
We may remedy this by using the Poisson summation formula
\begin{eqnarray}
F(\theta) \equiv \sum_J e^{-\Delta\tau UJ^2/2}e^{iJ\theta}
&=& \sum_{m=-\infty}^{\infty}\int_{-\infty}^{+\infty}dJ\,e^{2\pi iJm}
e^{-\Delta\tau UJ^2/2}e^{iJ\theta}\nonumber\\
&=&\sum_{m=-\infty}^{\infty} \sqrt{\frac{2\pi} {\Delta\tau U}}
e^{-\frac{1}{2\Delta\tau U}(\theta - 2\pi m)^2} \ .
\label{poisson}
\end{eqnarray}
This periodic sequence of narrow Gaussians is the Villain
approximation~\cite{villain} to the periodic function
\begin{equation}
F(\theta) \approx e^{K_\tau \cos(\theta)} \ ,
\label{eq:18}
\end{equation}
where 
we have dropped an irrelevant constant prefactor, and
\begin{equation}
K_\tau \equiv { 1 \over U \Delta\tau} \ .
\label{Ktau}
\end{equation}
Using this result in Eq.~(\ref{eq:16}) we finally arrive at the 
partition function of the anisotropic (2+1)D classical XY model
\begin{equation}
Z = \int{\cal D} \theta \exp\left\{\sum_{\langle
{\bf l,l^{\prime}}\rangle} K_{\langle {\bf l,l^{\prime}}\rangle}
\cos(\theta_{\bf l^{\prime}} - \theta_{\bf l})\right\} \ , 
\label{eq:19}
\end{equation}
where the sum is now over all near-neighbor bonds in both the space and time
directions, i.e.\ ${\bf l}=(x,y,\tau)$.  
For spatial bonds,
\begin{equation}
K_{\bf l,l^{\prime}} = K_x \ ,
\end{equation}
given by Eq.~(\ref{Kx}), while for temporal bonds
\begin{equation}
K_{\bf l,l^{\prime}} = K_\tau \ ,
\end{equation}
given by Eq.~(\ref{Ktau}).
It is implicitly assumed here that the
difference between the Villain action and the cosine term 
(which is small for small $\Delta\tau$) is in fact irrelevant
in the renormalization group sense.

Note that we need to take the limit $\Delta\tau \to 0$ which implies
$K_x \to 0$ and $K_\tau \to \infty$ such that the geometric mean,
\begin{equation}
K = \left(K_x K_\tau \right)^{1/2} = {t \over U} \ ,
\end{equation}
is finite.
Universality properties are unaffected~\cite{comment4} if we rescale 
space and time so that we obtain finally an {\it isotropic\/} (2+1)D XY model 
\begin{equation}
Z = \int{\cal D}\theta \exp\left\{ K \sum_{\langle 
{\bf l,l^{\prime}}\rangle}\cos(\theta_{\bf l^{\prime}} 
- \theta_{\bf l}) \right\}  \ .
\label{eq:isotropic}
\end{equation}
We are interested in the behavior of the boson Hubbard model at $T=0$,
which means taking the number of time slices, $M$, to infinity.
The coupling constant, $K$,
then controls the quantum rather than thermal
fluctuations~\cite{doniach,RANA-GIRVIN}.

Allowing for a non-integer boson density and/or including the random
potential in the boson Hubbard model, Eq.~(\ref{eq:bosonhubbard}),
makes the model more realistic but
complicates the situation by breaking the particle-hole symmetry
of the bosons.  This corresponds
to broken time-reversal symmetry for the quantum rotors 
(since particle number is represented by angular momentum) and hence leads to
{\em complex} weights in the corresponding classical statistical mechanical 
problem~\cite{fisher88a} which is no longer in the universality class of
the (2+1)D XY model.  The difficulty of complex weights
can be avoided by considering the
alternative approach to Eq.~(\ref{eq:16}) in which we integrate 
out the $\{\theta\}$ variables.  Let us do this first for the case of
integer boson density and no disorder. Adding the effects 
of disorder and non-integer density
will then be easy and will lead to a {\em real} action.

We first reexpress the cosine in Eq.~(\ref{eq:16}) as the best Villain
approximation to it, i.e.
\begin{equation}
\exp ( K_x \cos \theta ) \longrightarrow \sum_{m=-\infty}^\infty \exp \left\{
{1 \over 2 \tilde K_x} (\theta - 2\pi m )^2 \right\} \ .
\label{villainform}
\end{equation}
To determine $\tilde K_x$ we require that the
range of the functions on the two sides of Eq.~(\ref{villainform}) (as
$\theta$ varies from 0 to $\pi$) are the same (the precise angular
dependence of the two sides will be different but this is presumably
irrelevant). Using the Poisson summation formula, Eq.~(\ref{poisson})
and noting that $K_x \to 0$ from Eq.~(\ref{Kx}), one finds \cite{JKKN}
\begin{equation}
\tilde K_x = {1\over 2} \ln\left(2 \over K_x \right) \ .
\label{Ktildex}
\end{equation}
Inserting Eq.~(\ref{villainform}) into Eq.~(\ref{eq:16}) and Fourier
transforming~\cite{JKKN} one can carry 
out the $\{\theta\}$ integrations exactly. Their
effect is to enforce conservation of 
integer-valued currents defined by
\begin{equation}
{\bf J} = \left(J^x,J^y,J^\tau \right) \ .  
\end{equation}
In other words, the current should be divergenceless at every site in
space and time, i.e.\ it should obey a continuity equation
\begin{equation}
\partial_{\nu} J^{\nu} = 0 \ .
\label{eq:zerodiv}
\end{equation}
If $J_{(x,y,\tau)}^x$ lies on the bond between sites
$(x,y,\tau)$ and $(x+1,y,\tau)$ then it is convenient to define
$J_{(x,y,\tau)}^{-x}=-J_{(x-1,y,\tau)}$, etc. 
The divergence constraint 
is then imposed at each site by requiring that 
$\sum_{\nu} J_{({\bf r},\tau)}^{\nu} = 0$,
where $\nu$ runs over $\pm x, \pm y, \pm \tau$.
We thus obtain
\begin{equation}
Z \approx {\sum_{\{{\bf J}\}}}^{\prime} \exp\left\{ 
-{1\over 2} \sum_{({\bf r},\tau)} \sum_{\nu =
x,y,\tau}\tilde K_\nu\left(J^\nu_{({\bf r},\tau)}\right)^2\right\} \ , 
\label{eq:25}
\end{equation}
where the sum is over all integer values of the $J^{\nu}$ from
$-\infty$ to $\infty$,
the prime indicates the constraint that ${\bf J}$ be everywhere
divergenceless, and the couplings are
\begin{equation}
\tilde K_\tau = U \Delta\tau
\end{equation}
and $\tilde K_y = \tilde K_x$ given by Eq.~(\ref{Ktildex}).
Note that in taking the quantum limit, $\Delta\tau \to 0$, the spatial
couplings $\tilde K_x$ and $\tilde K_y$ diverge, while the coupling in
the time direction, $\tilde K_\tau$, tends to zero. This is the opposite
of what we found in the phase representation, see Eqs.~(\ref{Kx}) and
(\ref{Ktau}).

\begin{figure}[htb]
\centering
\epsfysize=14 cm
\leavevmode
\epsffile{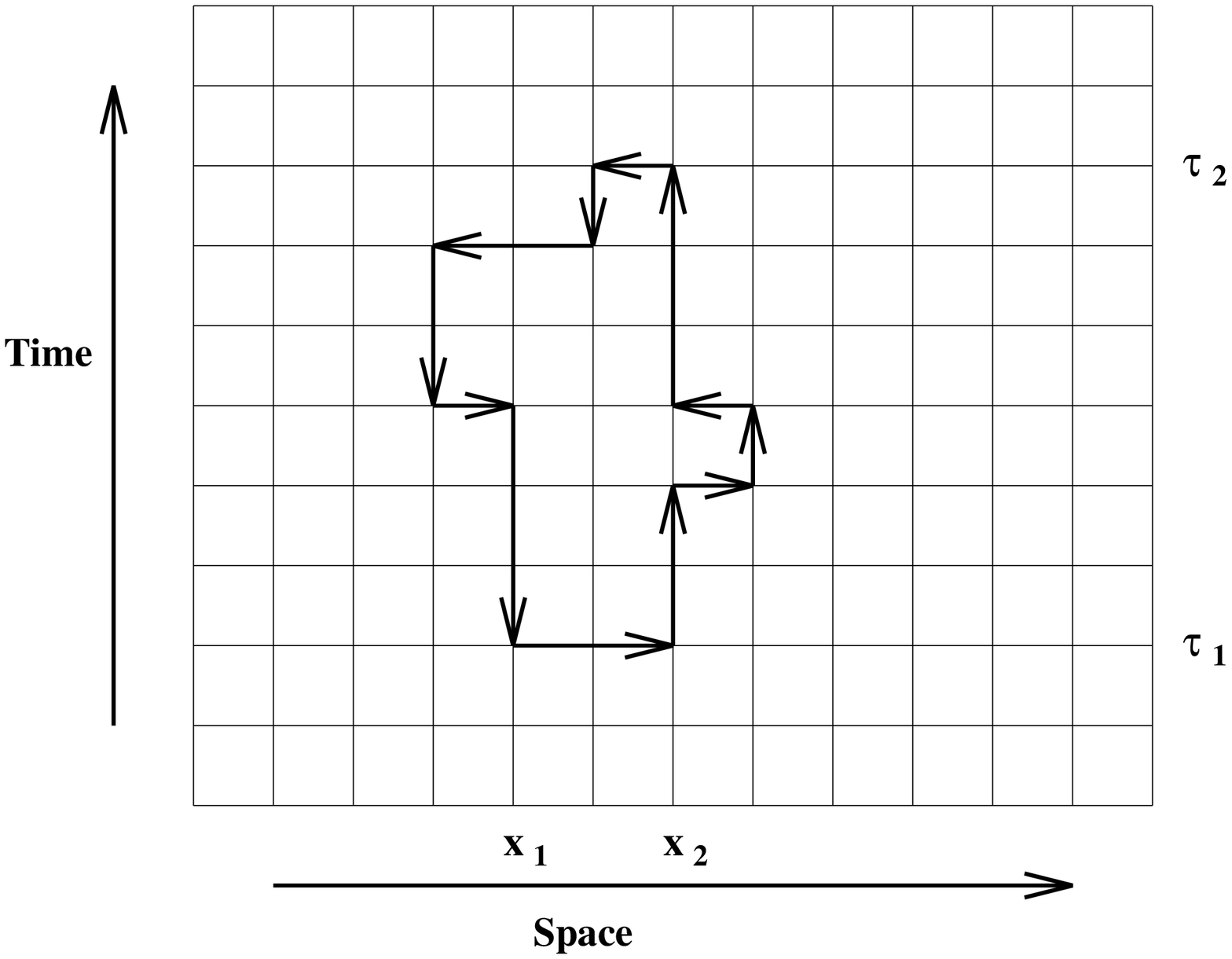}
\caption{
Typical closed loop of integer-values currents on the links of the space-time
lattice. A particle hops from $x_1$ to $x_2$ at time $\tau_1$, diffuses until 
time $\tau_2$ when it annihilates the hole it left behind.}
\label{fig:1}
\end{figure}

We interpret $J^\nu$ as the ``relativistic'' 
3-vector current with $(J^x,J^y)$
being the spatial current and $J^\tau$ being the particle density.  
Consider the divergenceless current configuration represented by 
the closed loop in the $x$-$\tau$ plane shown in Fig.~\ref{fig:1}.  
The physical interpretation of this is that at
time $\tau_1$ a boson hops from position $x_1$ to position $x_2$ 
creating an instantaneous burst of spatial current.  
This represents a tunneling event in which we assume the 
barrier is high enough that the tunneling time is small compared 
to the separation between time slices in our lattice and hence the
event can be treated as instantaneous.  
This approximation affects the ultraviolet details of the calculation 
but is irrelevant to the universal zero-frequency behavior.  
The two vertical lines represent time-like components of the 
current indicating that there is now a missing boson at $x_1$ 
and an excess of one boson at $x_2$.  
After some additional random motion, a boson hops back 
to the original site, leaving the system in the vacuum state at time
$\tau_2$.

This interpretation of the current is confirmed by consideration of the effect
of an external vector potential which modifies the potential energy of the
quantum rotors to
\begin{equation}
V = -t \sum_{({\bf r},\tau)}\sum_{\nu=x,y}
\cos\left(\theta_{({\bf r},\tau)} - \theta_{({\bf r}+\nu,\tau)} +
A^\nu_{\bf r}\right) \ , 
\label{eq:28}
\end{equation}
where $A^\nu_{\bf r}$ stands for the line integral of the vector 
potential along the
link from site ${\bf r}$ to its neighbor in the $\nu$-th direction.
Making this substitution modifies
Eq.~(\ref{eq:25})
with the result
\begin{equation}
Z \approx {\sum_{\{{\bf J}\}}}^{\prime} \exp\left\{ -
\sum_{({\bf r},\tau)} \left[\sum_{\nu =
x,y,\tau} {\tilde K_\nu \over 2}
\left(J^\nu_{({\bf r},\tau)}\right)^2
+ i\sum_{\nu=x,y}J^\nu_{({\bf r},\tau)}A^\nu_{({\bf r},\tau)}\right]
\right\} \ . 
\label{vectorpot}
\end{equation}
We note that
\begin{equation}
\left\langle J^\nu_{({\bf r},\tau)}\right\rangle = 
-i\frac{\delta \ln Z}{\delta A^\nu_{({\bf r},\tau)}} \ ;
\,\,\nu=x,y \ , 
\end{equation}
which means that $J^\nu_{\bf r}$ must thus be the full, physical, 
gauge-invariant current, not simply the paramagnetic piece of the current.

From our interpretation that $J^\tau$ is the particle density
it is now straightforward to include both disorder and
a value of the chemical potential which gives a non-integer density,
and one finds
\begin{equation}
Z = {\sum_{\bf J}}^{\prime} \exp\left\{-
\sum_{({\bf r},\tau)}\left[\sum_{\nu =
x,y,\tau} {\tilde K_\nu \over 2} \left(J^\nu_{({\bf r},\tau)}\right)^2
-\Delta\tau (\mu+v_{\bf r}) J^\tau_{({\bf r},\tau)}\right]
\right\} \ . 
\end{equation}
We now assume, as in Eq.~(\ref{eq:19}),
that the universality class is unchanged if we make the
couplings isotropic, i.e.\
\begin{equation}
Z = {\sum_{\bf J}}^{\prime} \exp\left\{- {1\over K}
\sum_{({\bf r},\tau)} \left[
{1 \over 2} {\bf J}^2_{({\bf r},\tau)}
- (\tilde\mu + \tilde v_{\bf r}) J^\tau_{({\bf r},\tau)}\right]
\right\} \ , 
\label{eq:pf}
\end{equation}
where
\begin{eqnarray}
\tilde \mu & = & {\mu \over U} \nonumber \\
\tilde v_{\bf r} & = & {v_{\bf r} \over U} \ ,
\end{eqnarray}
and $K$ is a dimensionless
coupling constant which has to be adjusted to bring the
system to the critical point. Varying $K$ corresponds to changing the
ratio $t/U$ in the boson Hubbard model, Eq.~(\ref{eq:bosonhubbard}),
keeping $\mu/U$ and $\Delta/U$ fixed.
Noting the invariance of the action under
\begin{eqnarray}
J^\tau &\longrightarrow& J^\tau + 1\nonumber\\
\tilde \mu + \tilde v &\longrightarrow& \tilde \mu + \tilde v + 1 \ ,
\end{eqnarray}
we take for simplicity the ``largest possible'' disorder by choosing
$\tilde \mu = 1/2, \tilde \Delta \equiv \Delta / U =1/2$.
The average particle density is then $1/2$. Note that this choice of
parameters has a statistical particle-hole
symmetry~\cite{comment5}
since, {\em upon ensemble
averaging} the Hamiltonian is invariant under the transformation $J^\tau
\to -J^\tau$, although the presence of the random potential destroys
microscopic particle-hole symmetry. We have argued above that lack
of {\em microscopic} particle-hole symmetry~\cite{comment5}
changes the universality class
from that of the (2+1)D XY model, so one can ask whether having
{\em statistical} particle-hole symmetry changes the universality class from
that of the generic Bose glass to superfluid transition. At
least in one dimension, the answer is no, as shown by
Fisher~\cite{fisherphysica},
and we shall assume that the same is true in $d=2$.

We have thus arrived at a representation of the original
quantum problem, involving integer link variables.
Noting that the $J^\tau$ represent the boson density, it can be
thought of as an imaginary time
``world-line'' path-integral representation of the problem, simplified
to the extent that it treats just the phase fluctuations of the
underlying Hamiltonian.

The partition function, Eq.~(\ref{eq:pf}) can be written in terms of
an effective (2+1)D classical Hamiltonian or action, given by
\begin{equation}
H_{\text{V}}=
{1\over K}
\sum_{({\bf r},\tau)} \left[
{1 \over 2} {\bf J}^2_{({\bf r},\tau)}
- (\tilde\mu + \tilde v_{\bf r}) J^\tau_{({\bf r},\tau)}\right] \ .
\label{eq:nocoul}
\end{equation}
Evidently, when
$\tilde v_{\bf r}=0$,
integer values of $\tilde \mu$ can be absorbed into the definition
of $J^\tau_{({\bf r},\tau)}$, so the model reduces to the (2+1)D
Villain model, which is in the
same universality class as the (2+1)D XY model. These points, are, however,
just special multicritical points and the generic behavior is {\em not} that
of the  XY model~\cite{fisher89c}.

We have already noted that the time component, 
$J^\tau_{({\bf r},\tau)}$, 
of the link variables corresponds to the particle
density or boson occupation number. 
Long-range Coulomb forces can then be introduced
in the following way.
\begin{eqnarray}
\label{eq:ourH}
H &=& H_V+H_{\text{C}}\\
H_{\text{C}} &=& 
\frac{{e^{\ast}}^2}{K}\sum_{\tau}\sum_{\langle {\bf r},{\bf r}^{\prime}
\rangle }
(J_{({\bf r},\tau)}^{\tau}-n_0)G({\bf r}-{\bf r}^{\prime})
(J_{({\bf r}^{\prime},\tau)}^{\tau}-n_0) \ .
\end{eqnarray}
Here $e^{\ast}$ is the effective boson charge,
and $n_0$, which represents the compensating background charge,
is the average particle density,
and $G$ is the Coulomb interaction. 
In our simulations with long-range interactions,
the particle number was always kept constant,
as opposed to the case where only short-range interactions were present
where we always allowed the particle number to fluctuate.
Calculations of the
Coulomb interaction, $G({\bf r})$ must allow for
the finite lattice size and periodic boundary conditions.
We do this by the usual Ewald method~\cite{totsuji,fisher79}.
Another way is by a lattice Green's function:
\begin{equation}
G({\bf r})=\frac{2\pi}{L^2}\sum_{{\bf k}\neq 0}
\frac{\cos ({\bf k \cdot r})}
{[4 - 2 \cos(k_x) - 2 \cos(k_y)]^{1/2}} \ ,
\label{eq:green}
\end{equation}
where ${\bf k} = (2\pi/L)(n_x, n_y)$, with
$n_x,n_y=0, \dots, L-1$,
The term with ${\bf k}=0$ is removed to ensure charge neutrality.
For large distances and large lattices the Ewald sum and the
lattice Green's function become almost identical and approach $1/r$. 
However, close to the origin the two forms are somewhat different. 
If the critical properties are universal they should not depend
on the specific form of the potential close to the origin.
We use this as a test of our computer codes
and of the universality of our results.
Indeed as we shall see the two forms yield equivalent results.

\section{Scaling Theory}\label{sec:scaling}

In order to better understand the universal features
of the phase transition it is very useful to
consider the scaling behavior of various physical quantities
in the regime of the diverging correlation length.
Such considerations not only tell us why the conductivity
is universal but will tell us how to analyze 
experimental and Monte Carlo data to determine that one
is actually in the critical (scaling) regime.

From now on, we shall denote the number of time slices by
$L_\tau$, rather than $M$, so the space-time lattice is of size
$L \times L \times L_\tau$. Periodic boundary conditions will be
applied. Note that the ground state {\em energy}
density of the original 2D quantum problem is related to the 
{\em free energy} density of the (2+1)D equivalent classical problem since
\begin{equation}
f = -\lim_{T \to 0} {k_B T \over (aL)^2} \ln Z = 
-{\hbar \over V} \ln \mbox{Tr} e^{-H} \ ,
\label{fedensity}
\end{equation}
where $H$ is given by Eq.~(\ref{eq:ourH}),
$V = (aL)^2 L_\tau \Delta\tau \hbar$ is the ``volume'' of the
(2+1) D space-time system, with $a$ the lattice spacing in
the spatial directions and $\hbar \Delta\tau$ the
lattice spacing in the (imaginary) time direction.

Since space and time are not equivalent we have two correlation
``lengths'', $\xi$ in the space direction and $\xi_\tau$ in the time
direction. These two correlation lengths will diverge with different
exponents at the critical point and we can define the dynamical
exponent, $z$, through the relation
\begin{equation}
\xi \sim \delta^{-\nu},\ \ \ \xi_\tau \sim \xi^z \ ,
\end{equation}
where $\delta$ measures the distance from the critical point, $K_c$,
i.e.\ $\delta = (K - K_c) / K_c$.
There is a microscopic frequency, $\omega_c$, related to the 
lattice spacing $\hbar\Delta\tau$,
in the time direction by
$ \omega_c = 2 \pi /( \hbar \Delta\tau), $
so we can relate $\xi_\tau$ more precisely to $\xi$ as
\begin{equation}
\xi_\tau= {1 \over \omega_c} \left( \frac{\xi}{b} \right)^z \ ,
\label{omegac}
\end{equation}
where $b$ is a microscopic length of order the lattice spacing, $a$.

\subsection{Stiffness}

First we discuss the scaling theory describing the singular behavior of
the free energy density near the critical point. 
From Eq.~(\ref{fedensity}) one sees that $f/\hbar$ has dimensions of inverse
(length$^d$ $\times$ time).
Hyperscaling~\cite{mef} states that multiplying
the singular part of this free energy density, $f_s$,
by the (2+1)D correlation volume, $\xi^d \xi_\tau$,
one obtains a constant, $A$ say, of order unity
as the critical point is approached, i.e.
\begin{equation}
{f_s \over \hbar} \xi^d \xi_\tau = A \ .
\end{equation}
In this section, we will
frequently give results for arbitrary space
dimension, $d$, even though we are ultimately interested in the case of $d =
2$. One can consider $A$ to be a critical amplitude for a {\em dimensionless}
quantity (or combination of quantities)
which is finite at criticality.
According to two-scale factor 
universality~\cite{stauffer,aharony74,hohenberg76,aharony76,weichman91},
such quantities are not only constants, but are also {\em
universal}.
 
We next discuss the scaling of the extra free energy cost to impose a
twist on the phase of the condensate. We will use this to
locate the critical point, $K_c$, to high accuracy. 
The extra free energy density is related to the superfluid stiffness, also
called the helicity modulus~\cite{helicity}, which is proportional to
the superfluid density of the system.
A uniform twist in the phase of the order parameter can be introduced 
by applying a twist of size $\Theta$ at the boundary, 
in (say) the $x$ direction. 
This will then give rise to a phase gradient 
\begin{equation}
\nabla \theta=\Theta/(aL) \ .
\end{equation} 
The (zero-frequency) stiffness,
$\rho$, is then defined by~\cite{helicity,hertz85,cha91}
\begin{equation}
{\delta f_s \over \hbar}=\frac{1}{2}\rho (\nabla \theta)^2  \ ,
\label{eq:stiffluc}
\end{equation}
so
\begin{equation} 
\rho = {(aL)^2 \over \hbar} { \partial^2 f_s \over \partial \Theta^2} \ .
\label{eq:31}
\end{equation}
Since $\Theta$ is dimensionless, $\rho $ has dimensions of
inverse (length$^{d-2}$ $\times$ time). Hence using hyperscaling and 
two-scale factor universality we obtain
\begin{equation}
\rho  \xi^{d-2} \xi_\tau = C \ ,
\label{rho:univ}
\end{equation}
where $C$ is another universal constant. Consequently,
\begin{equation}
\rho \sim \xi^{-(d + z - 2)} \ ,
\label{rho:scale}
\end{equation}
which is a generalization of the Josephson scaling relation for the
classical transition, $\rho_s \equiv (m / \hbar)^2 \hbar \rho \sim 
\xi^{d - 2}$. The difference is that $d$ is replaced by $d+z$ for the
quantum transition. This replacement also holds for other hyperscaling
relations (i.e.\ those scaling relations involving the space dimensionality).

\subsection{Conductivity}\label{subsec:cond}
We can extend the notion of a superfluid stiffness, $\rho$,
to a frequency-dependent stiffness, $\rho(i\omega_n)$, where
$\omega_n = 2 \pi n k_B T / \hbar$ is the Matsubara frequency.
The conductivity is then related to $\rho(i\omega_n)$ 
by the Kubo formula~\cite{cha91}
\begin{equation}
\sigma(i\omega_n) = 2\pi G_Q
\frac{\rho(i\omega_n)}{\omega_n} \ ,
\label{eq:kubo}
\end{equation}
where $G_Q = R_Q^{-1}$, with $R_Q$ defined in Eq.~(\ref{rq}), is
the quantum of conductance. The quantity $\rho$ defined
in  the previous section is given by $\rho \equiv \rho(0)$. We emphasize
that $\rho$ is the stiffness and not the resistivity.
Close to the critical point we can generalize Eq.~(\ref{rho:univ})
to finite frequency~\cite{fisher90a} by the following scaling
assumption
\begin{equation}
\rho(i\omega_n) 
= \xi^{2-d}\xi_{\tau}^{-1}\widetilde\rho(\omega \xi_\tau) \ .
\label{eq:rhoomega}
\end{equation}
Since the argument of the scaling function $\widetilde\rho$ is
dimensionless, and so has no non-universal metric factors associated
with it, the {\em entire scaling function} $\widetilde\rho(x)$ is universal.
Clearly,
$\widetilde\rho(0) = C$, the same universal constant that appears
in Eq.~(\ref{rho:univ}). Furthermore, since
$\rho(i\omega_n)$ is finite at finite frequency even at the critical
point, one must have, for large $x$, the asymptotic behavior
\begin{equation}
\widetilde\rho(x) = D x^{(d+z -2)/z} \ ,
\end{equation}
where $D$ is again universal, in order that the dependence on $\xi$
and $\xi_\tau$ cancels at criticality.
Substituting this into Eq.~(\ref{eq:kubo}) and noting
Eq.~(\ref{omegac}), one has, at criticality,
\begin{equation}
\sigma^\ast \equiv
\lim_{\omega_n\rightarrow 0}\sigma(i\omega_n)=2\pi D\, \sigma_Q\,b^{2-d}
\left (\frac{\omega}{\omega_c} \right)^{(d - 2)/z} \ .
\label{sigma:crit}
\end{equation}
Immediately we see that when $d=2$ all microscopic lengths and
frequencies
drop out so the d.c.\ conductivity is universal~\cite{fisher90a}
at the critical point,
given only by fundamental constants and the universal
dimensionless number $D$. The universality of the d.c.\ conductivity is
analogous to the universal jump~\cite{NK} in
$(\hbar/ m)^2 \rho_s / k_BT_c$
at the finite-temperature Kosterlitz transition~\cite{KT}.  In fact
this quantity corresponds, essentially, to Eq.~(\ref{eq:kubo})
with  $\hbar \omega_n$ replaced by $ k_B T_c$.

Strictly speaking Eq.~(\ref{sigma:crit}) only refers to the singular part of
the conductivity. However since we approach an insulating
phase where the conductivity must be zero, the
conductivity cannot have an analytic part at the critical
point.

\subsection{Compressibility}

The compressibility, $\kappa$, is defined by
\begin{equation}
\kappa = {\partial n \over \partial \mu} =
{\partial^2 f \over \partial \mu^2} \ ,
\label{compress}
\end{equation}
where $n$ is the boson density and $\mu$ the chemical potential.
We can write an expression equivalent to Eq.~(\ref{eq:stiffluc}) for
the compressibility by noting~\cite{fisher89c} that the Josephson
relation (for imaginary time) is $\delta \mu = \partial \theta /
\partial \tau$, so 
\begin{equation}
\delta f = \frac{1}{2}\kappa (\partial_\tau \theta)^2  \ ,
\label{eq:kappafluc}
\end{equation}
i.e.\ we apply a twist 
in the (imaginary) time direction instead of along one of the space directions.
Note that it is the {\em total} compressibility which enters this
expression~\cite{fisher89c}. Following the arguments that led to 
Eqs.~(\ref{rho:univ}) and (\ref{rho:scale}) one finds
\begin{equation}
{ \kappa \over \hbar} \xi^d \xi_\tau^{-1}
\equiv { \kappa \over \hbar} \xi^d \omega_c \left( b \over \xi \right)^z
= \mbox{const.}
\label{kappa:univ}
\end{equation}
so
\begin{equation}
\kappa \sim \xi^{-(d - z)} \ .
\label{kappa:exponent}
\end{equation}
Fisher et al.~\cite{fisher89c} have argued that $z = d$ at the Bose glass
to superfluid transition and so the
compressibility is finite at criticality. We shall see that our
numerical results support this. Note that even in this case, the
compressibility is non-universal at criticality, because the
non-universal factors $b$ and $\omega_c$ appear in 
Eq.~(\ref{kappa:univ}).
One can also determine the form of the wave-vector dependent
compressibility at criticality, following the scaling arguments that we
used above to determine the conductivity. One finds
\begin{equation}
\kappa(k) \sim k^{d-z} \ .
\label{eq:kk}
\end{equation}

\section{Quantities of Interest and Finite Size Scaling}\label{sec:fss}
 
In this section we show how to calculate the quantities of interest from
the Monte Carlo simulations, and we discuss the finite size scaling techniques
which we will need.
Having demonstrated in the last section, that the lattice spacings, $a$
and $\hbar\Delta\tau$, do not enter expressions for universal quantities,
such as the conductivity at the critical point, we set these lattice
spacings (and $\hbar$) to unity from now on.

To perform the quenched disorder averages it is necessary to 
first do a ``thermal'' average over the $J^{\nu}$ variables,
denoted by $\langle \cdots\rangle $,
for a fixed realization of the quenched disorder potential,
and then average observables
over the quenched disorder $v_{\bf r}$, 
which we indicate by $[\cdots]_{\text{av}}$.

\subsection{Stiffness}\label{subsec:stiff}
To calculate the uniform stiffness, $\rho(0)$,
note from Eq.~(\ref{eq:28}) that a uniform
twist in the $x$ direction becomes equivalent to considering the system in
the presence of an external vector potential of the form
\begin{equation}
A^x_{\bf r} = \partial_x \theta \delta_{x,\nu} \ .
\label{eq:30}
\end{equation}
From Eqs.~(\ref{eq:28}), (\ref{eq:31}) and (\ref{eq:30}) one finds that
\begin{equation}
\rho(0) = \frac{1}{L^d L_\tau} 
\left [ \left\langle\left(\sum_{({\bf r},\tau)} 
J^x_{({\bf r},\tau)}\right)^2\right\rangle \right ]_{\text{av}} \ .
\end{equation}

Near the critical point, the correlation length is much larger than the
size of the system, so finite-size effects will be important.
We  therefore need to derive a finite-size scaling form~\cite{privman}
for the stiffness. The basic finite-size scaling hypothesis is that the
size of the system only appears in the ratio $L/\xi$, and, for quantum
problems, the corresponding ratio in the time direction,
$L_\tau/\xi_\tau$. Thus we have
\begin{equation}
\rho(0) = \xi^{-(d+z - 2)} P(L/\xi,\ L_\tau/\xi_\tau) \ ,
\end{equation}
which can be more conveniently expressed as
\begin{equation}
\rho(0) = {1 \over L^{d+z - 2}} \bar\rho \left(
L^{1/\nu} \delta,\ {L_\tau \over L^z} \right) \ ,
\label{eq:rfinscale}
\end{equation}
where $P$ and $\bar \rho$ are scaling functions.
It is thus essential to work with system shapes for which
the aspect ratio
\begin{equation}
c =  L_\tau/L^{z}
\label{aspect}
\end{equation}
is a constant, otherwise the scaling function $\bar \rho$ depends on
two variables and is complicated to analyze.
If this is done, $L^{d+z-2}\rho$ is {\em independent} of $L$
at the critical point $\delta = 0$.
Furthermore, in 
the disordered state, the system is insensitive to changes in the
boundary conditions if the size is bigger than the correlation length,
so $L^{d+z-2} \rho$ will
{\em decrease} (exponentially) with increasing $L$. By contrast, in the
ordered state, $\rho$ tends to a constant so $L^{d+z-2} \rho$
{\em increases} with increasing $L$. Thus,
the critical point is located at the intersection of curves for
$L^{d+z-2} \rho$ as a function of coupling $K$ for different lattice
sizes. One can then determine $\nu$ from Eq.~(\ref{eq:rfinscale}) by
requiring that the data for different sizes (but fixed aspect ratio)
collapse on top of each other in
a plot of $\rho(0)L^{d+z-2}$ against $L^{1/\nu}\delta$. Note that in order
to choose the sample shapes in the simulation, 
we need (unfortunately) to have
already made a choice for $z$.

Since the current is divergenceless, we can divide the configurations 
into different topological classes according to the winding number of 
the boson world lines around the torus of size $L$ in the space direction
\begin{equation}
n_x \equiv L^{-1} \sum_{({\bf r},\tau)} J^x_{({\bf r},\tau)} \ ,
\label{eq:32}
\end{equation}
so that the stiffness is simply proportional to the mean-square winding number
\begin{equation}
\rho(0) = \frac{1}{L_\tau} \left [ \left\langle n^2_x
\right\rangle \right ]_{\text{av}} \ .
\label{eq:33}
\end{equation}
It is instructive to comment on the analogy to the Feynman ring exchange 
picture of superfluidity in liquid helium~\cite{feynman}.
Rather than viewing a nonzero winding of a boson world line as an event 
involving a single boson, we can view it as formed by adding up a chain 
of hops of many bosons.
This is closely analogous to a Feynman ring exchange,
and adds some perspective on the transition in our model.
The superfluid stiffness arises due to macroscopic condensation
of ring exchanges of global currents carrying nonzero winding number,
and the critical point is where the gain in free energy
from the ``entropy'' of the ring exchanges matches their energy cost.
In the presence of macroscopic ring exchanges which wind around the sample,
the free energy is sensitive to Aharonov-Bohm flux (boundary condition twists)
and hence the system exhibits off-diagonal long-range order in the conjugate
phase variable. 

\subsection{Conductivity}
The frequency-dependent stiffness involves
the Fourier transform of the current-current correlation function
\begin{equation}
\rho(i\omega_n) = \frac{1}{L^2L_{\tau}} [\langle  | \sum_{({\bf r},\tau)}
e^{i\omega_n\tau}
J^x_{({\bf r},\tau)} |^2 \rangle ]_{\text{av}} \ ,
\label{eq:rhon}
\end{equation}
where, with the lattice spacings in the space and time
directions set to unity, the Matsubara frequency is given by
$\omega_n = 2\pi n/L_\tau$, and $\tau$ is now an {\em integer},
$1 \le \tau \le L_\tau$, denoting a particular time slice. In these units,
the conductivity is still given by Eq.~(\ref{eq:kubo}).

\subsection{Compressibility}

From Eq.~(\ref{compress}) it follows that the zero wave vector
compressibility is given by
\begin{equation}
\kappa(0) = \frac{1}{L^2L_{\tau}}[\langle N_b ^2\rangle -\langle
N_b\rangle ^2]_{\text{av}} \ ,
\label{eq:kappa}
\end{equation}
where $N_b$ is the total number of particles,
\begin{equation}
N_b = {1\over L_\tau} \sum_{({\bf r},\tau)}J_{({\bf r},\tau)}^{\tau} \ .
\end{equation}
The last term in Eq.~(\ref{eq:kappa}) 
involves the square of a
thermal average. This term is thus likely to give systematic
errors~\cite{young88}
if determined within one replica, so we evaluate it as
$[ \langle N_{\alpha}\rangle \langle N_{\beta}\rangle ]_{\text{av}}$,
where the indices refer to two different replicas.

If global moves are not included, the boson density is a constant,
and consequently $\kappa$, as defined, is zero. However,
the wave-vector dependent compressibility
$\kappa(k)$ is nonzero, and one can obtain~\cite{batrouni90},
estimates of
$\kappa=\kappa(0)$ by taking the limit $k \rightarrow 0$, even when
global moves are not performed.

The finite-size scaling form for the compressibility follows from
arguments similar to those used above for the stiffness and is
\begin{equation}
\kappa= {1 \over L^{d-z} } \widetilde\kappa
\left( L^{1/\nu}\delta,\ \frac{L_\tau}{L^z} \right) \ .
\label{eq:compfin}
\end{equation}

\subsection{Correlation functions}\label{subsec:cf}

Consider the following correlation function
\begin{equation}
C({\bf r,r'},\tau,\tau') =  [\langle e^{i(\hat\theta_{\bf r}(\tau)-
\hat\theta_{\bf r'}(\tau'))}\rangle ]_{\text{av}} \ ,
\end{equation}
where the $\hat\theta$'s are operators for for the phase
of the bosons, and  $e^{i\hat\theta_{\bf r}(\tau)} =
e^{\tau H}e^{i\hat\theta_{\bf r}}e^{-\tau H}$. We shall see that 
this correlation function  
gives information on a third critical exponent, $\eta$,
defined in Eq.~(\ref{eq:crscale}) below, 
in addition to the exponents $\nu$ and
$z$ already discussed.
Due to translational invariance,
$C({\bf r,r'},\tau,\tau')=C({\bf r-r'},\tau-\tau')$.

We shall here only consider two basic types of correlations.
The equal-time correlation function where $\tau=0,{\bf r}=(x,0)$, and the
time-dependent correlation function at time $\tau$ and ${\bf r}=0$.
By redoing the argument which led from 
Eqs.~(\ref{hqr}) to (\ref{eq:25}) for the
correlation function rather than for the partition function, one finds that
the equal-time correlation function can be expressed as 
\begin{equation}
C_x(r) = [\ \langle \prod_{{\bf r}\; \rm \in path}
\exp \left\{ -\frac{1}{K} (J_{( {\bf r},\tau)}^\nu+\frac{1}{2}) \right\}
\ \rangle\  ]_{\text{av}} \ ,
\label{eq:cr}
\end{equation}
where ``path'' is any path on the lattice at fixed $\tau$
connecting two points a distance $r$ apart
along the $x$-direction. For each link on the
path $\nu = x$ or $y$,  depending on whether the link
is along the $x$ or $y$ direction. The simplest case, which was
used in the simulations, is the straight line path, in which case all
the link variables in Eq.~(\ref{eq:cr}) are $J^x$.
A very similar result is is found for the usual Villain model~\cite{JKKN}.
Since the Hamiltonian is
invariant under a sign change of $J_x$, it follows that
$C_x(r)=C_x(-r)$.

We now turn to the time-dependent correlation function,
\begin{equation}
C_+(\tau) \equiv C({\bf r} = 0, \tau) =
[\langle e^{i(\hat\theta_{\bf r}(\tau)-
\hat\theta_{\bf r}(0))}\rangle ]_{\text{av}} \ . 
\label{cplustau}
\end{equation}
Physically this is the Green's function for creating a
particle at imaginary time 0 and destroying it at time $\tau$.
This correlation function can be expressed in terms of the link variables
in the following form
\begin{eqnarray}
C_+(\tau)& = &
\Bigl[ \Bigl\langle
\prod_{\tau \rm\; \in path} 
\exp
\Biggl\{ -\frac{1}{K}
(\frac{1}{2}+J_{({\bf r},\tau)}^{\tau} -\tilde\mu_{\bf r}) \nonumber\\
& &  -\frac{{e^{\ast}}^2}{K}
\left[{\sum_{{\bf r}}^{\prime}}
(J_{({\bf r}^{\prime},\tau)}^{\tau}-n_0)
G({\bf r}-{\bf r}^\prime) +
\frac{1}{2L^2}\sum_{{\bf r}}(G(0)-G({\bf r}))\right]
\Biggr\}\ 
\Bigr\rangle \ \Bigr]_{\text{av}} \ .
\label{eq:ct+}
\end{eqnarray}
In this expression, ``path'' is the straight
line path between two points with the same space coordinate,
${\bf r}$,
starting at imaginary time equal to 0, say, and ending at a later time
$\tau$. A more general expression for a path wandering in
the space directions can also be derived.

One can also consider
\begin{equation}
C_-(\tau) =  [\langle e^{-i(\hat\theta_{\bf r}(\tau)-
\hat\theta_{\bf r}(0))}\rangle ]_{\text{av}} \ , 
\label{cminustau}
\end{equation}
which is the Green's function for creating a {\em hole}
at imaginary time 0 and destroying it at $\tau$.
In terms of the link variables
\begin{eqnarray}
C_-(\tau)&=&
\Bigl[ \Bigl\langle
\prod_{\tau \rm\; \in path} 
\exp
\Biggl\{ -\frac{1}{K}
(\frac{1}{2}-J_{({\bf r},\tau)}^{\tau} -\tilde\mu_{\bf r}) \nonumber\\
& & +\frac{{e^{\ast}}^2}{K}
\left[{\sum_{{\bf r}}^{\prime}}
(J_{({\bf r}^{\prime},\tau)}^{\tau}-n_0)
G({\bf r}-{\bf r}^\prime) -
\frac{1}{2L^2}\sum_{{\bf r}}(G(0)-G({\bf r}))\right]
\Biggr\} \ 
\Bigr\rangle \  \Bigr]_{\text{av}} \ .
\label{eq:ct-}
\end{eqnarray}
Except when there is statistical particle-hole
symmetry~\cite{comment5},
$C_+(\tau)\neq C_-(\tau)$~\cite{comment2}. However, one can show that
$C_+(\tau)=C_-(L_\tau-\tau)$, which corresponds to the equivalence
between a particle traveling forwards in time and a hole traveling
backwards. This will be useful in the simulation
because the statistics get worse with increasing $\tau$ so, for $\tau >
L_\tau / 2$, it is better to compute $C_-(L_\tau - \tau)$ than
$C_+(\tau)$. As required, the correlation functions are periodic, i.e.
$C(\tau)=C(\tau+L_\tau)$ for both $C_-$ and $C_+$.

Following Ref.~\onlinecite{fisher89c} we now make the assumption that the
long-distance, large-time behavior of the correlation functions will be
given by the scaling form
\begin{equation}
C({\bf r},\tau)=r^{-(d + z -2+\eta)}f(r/\xi,\tau/\xi^z) \ ,
\label{eq:crscale}
\end{equation}
which defines the exponent $\eta$.
If ${\bf r}$ approaches
zero but $\tau$ remains finite, the correlation functions must remain
finite and nonzero. Thus we obtain
\begin{equation}
C({\bf r = 0},\tau) = \tau^{-(d + z -2+\eta)/z}g(\tau/\xi^z) \ .
\label{eq:ctscale}
\end{equation}
At the critical point we should therefore have
\begin{eqnarray}
C_x(r) &\sim& r^{-y_x}\nonumber\\
C_\tau(\tau) &\sim& \tau^{-y_\tau} \ , \nonumber\\
\end{eqnarray}
where
\begin{eqnarray}
y_x  &=& d + z-2 +\eta\nonumber\\
y_\tau &=& (d + z-2 +\eta)/z \ .
\label{eq:expo}
\end{eqnarray}
Thus the power law fall-off of the correlation functions at criticality
determines both $\eta$ and $z$ provided the correlation functions
can be evaluated for large enough
system sizes that finite-size corrections are unimportant.

\section{Monte Carlo Methods}\label{sec:methods}

\begin{figure}[htb]
\centering
\epsfysize=14 cm
\leavevmode
\epsffile{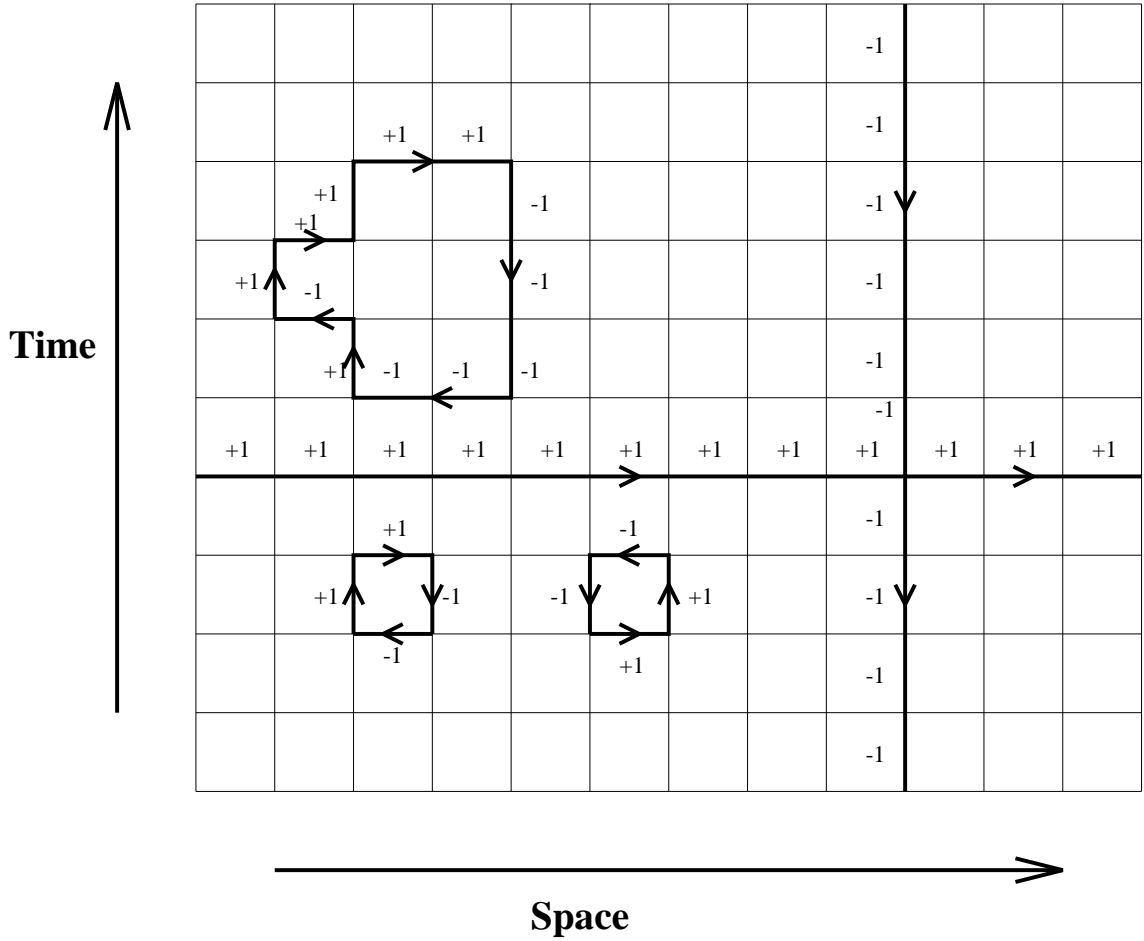}
\caption{
Schematic picture of a local move (lower left corner), 
and global move (right side of the picture).
In the local moves
changes of $\pm1$ are attempted in the currents circulating 
around a plaquette.
In the global the current is changed by
$\pm1$ along straight lines across the system,
thereby changing the winding number, if the line is along a
space direction, or the boson number if the line is in the time
direction.
The numbers indicate how the link variables are changed. 
Many local
moves will change the current around an arbitrary loop as indicated
in the upper left corner.
A global move which destroys a boson is shown, and 
also another global move which 
increases the winding number along the $x$ direction by one.}
\label{fig:linkcur}
\end{figure}

To satisfy the zero divergence criterion in Eq.~(\ref{eq:zerodiv}) our
basic (local) Monte Carlo move consists of 
changing all the link variables around one plaquette
simultaneously in the manner shown in the lower left corner of the
Fig.~\ref{fig:linkcur}, thus
changing the local current. Two of the link variables are increased
by one, the other two decreased by one. An equivalent move going
in the other direction is also used, i.e.\ the plusses and minuses
are interchanged.
In addition, we need to include non-local moves to
fully equilibrate the system.
The global moves consist of changing by $\pm 1$
a line of link variables stretching all through the system.
Nonlocal moves are included in all three directions $\delta=x,y,\tau$,
except when the model has long-range interactions, in
which case no global moves in the time direction are performed in
order to keep the particle number constant.
Global moves in the time direction amount to either introducing or
destroying a boson.
It is easy to see that global moves in the space directions
correspond to a change
in the winding number~\cite{batrouni90,scalettar91}, defined in 
Eq.~(\ref{eq:32}).
The nonlocal moves we use are illustrated in
Fig.~\ref{fig:linkcur}. One Monte Carlo sweep of the lattice consists
of a sweep of local moves followed by a sweep of global moves.

Due to the continuity equation, Eq.~(\ref{eq:zerodiv}), 
the sum of $J_{\bf r}^{\tau}$ at a given time slice,
$\sum_{\bf r} J_{\bf r}^{\tau}$,
is always the same for any value of $\tau$, albeit this constant may
vary as a function of Monte Carlo time because of global
moves in the time direction. 
Likewise, the sum of $J^x$ in any $y$-$\tau$ plane will be the
same for all such planes at a fixed Monte Carlo time, and similarly
for the sum of the $J^y$.

Expectation values of observables have to be computed by 
quenched disorder averaging, 
which is known from the study of spin glasses~\cite{young86}
to have many potential pitfalls.
Close to the critical point we typically have to average over from 
200 to 1000 different realizations of the disorder, and somewhat 
fewer away from the critical point. 
It is crucial to carefully assure
that the $J^{\nu}$ variables are thermally equilibrated.
The equilibration time at the critical point  for our update scheme varies
with system size $L$ as
$\tau_{\rm mc} \sim L^{z_{\rm mc}}$, where $z_{\rm mc}$ is the Monte
Carlo dynamic exponent.
For the short-range interaction case we have determined~\cite{MW-SMG}
$z_{\rm mc} \approx 6$ so that extreme caution is required in attempting to
equilibrate large lattices.
We take an approach similar to what has been done for spin glass
systems~\cite{young88}. 
Two identical replicas are run in parallel for a given realization 
of the disorder. 
We define the ``Hamming'' distance between replicas $\alpha$ 
and $\beta$ as:
\begin{equation}
h_{\alpha,\beta}^{\nu}(t) 
= \sum_{( {\bf r},\tau)}\left[ J^{\nu}_{ ( {\bf r},\tau),\alpha} (t_0 + t)
-J^{\nu}_{({\bf r},\tau),\beta} (t_0 + t) \right]^2 \ ,
\end{equation}
where $t_0$ is the number of Monte Carlo sweeps (MCS) used for equilibration,
and $t$ is the number of subsequent MCS.
We also define a ``Hamming'' distance for one replica 
at two different Monte Carlo times,
\begin{equation}
h_{\alpha}^{\nu}(t) 
= \sum_{( {\bf r},\tau)}\left[ J^{\nu}_{ ( {\bf r},\tau),\alpha}(t+t_0)
-J^{\nu}_{({\bf r},\tau),\alpha}(t_0) \right]^2 \ .
\end{equation}
We determine  $[ h_{\alpha,\beta}^{\nu}(t_0) ]_{\text{av}}$ and
$[h_{\alpha}^{\nu}(t_0)]_{\text{av}}$ for a sequence 
of values of $t_0$ increasing
exponentially, $t_0=10,30,100,300,1000, \dots$, up to $t_0 = T_0$,
so $T_0$ is both
the number of MCS for measurement and the number of MCS
for equilibration.
If $t_0$ is sufficiently large that the system has equilibrated,
one has
$[ h_{\alpha,\beta}^{\nu}(t_0) ]_{\text{av}} = 
[h_{\alpha}^{\nu}(t_0)]_{\text{av}}$,
and we made sure that this condition was fulfilled, at least for $t_0 =
T_0$. 
To achieve equilibration we took
$T_0$ to be of order 3,000 for the smaller
system sizes but found that we needed up to 30,000 for the larger sizes. 
Since the different disorder realizations give statistically 
independent thermal averages, 
we can estimate the statistical error from the standard deviation
of the results for different samples. Note that there are big sample to
sample fluctuations, so it is necessary to average over a large number
of samples.
In order to study as many samples as possible within the available
computer time, we only run each sample for the minimum number of MCS
necessary to get a few statistically independent measurements.
This is why
the number of sweeps for averaging is the same as the number
used for equilibration.

\section{Short range interactions}\label{sec:dirtyb}

In this section we shall 
assume that no long-range Coulomb interactions
are present.
Furthermore, we shall always take the
random chemical potential to be specified by
\begin{equation}
\tilde\mu=\frac{1}{2},\;\;\tilde\Delta=\frac{1}{2} \ .
\end{equation}
The reason for this choice of $\tilde\mu$ is that we want to be as far
away as possible from any Mott insulator phase~\cite{fisher89c}, and
these are centered on integer values of $\tilde\mu$ for weak disorder.
The choice of $\tilde\Delta$ was influenced by the need to make the disorder
not too small (otherwise the effects of disorder would only be seen for
large sizes which we are unable to simulate) and also not too large,
because this effectively makes $U$ small, and so, again, the asymptotic
behavior may only set in for large sizes. In the absence of more detailed
information, it seems sensible to make all the important couplings of
comparable size.
We again emphasize that universal quantities like the critical 
conductivity are independent of these details.

Some inequalities involving the critical exponents, $\nu,\ \eta,$ and
$z$ have been obtained. First of all, Fisher et al.~\cite{fisher89c}
have argued that the compressibility is finite at the transition and so
\begin{equation}
z = d \ .
\end{equation}
Fisher et al.~\cite{fisher89c} also argue that
\begin{equation}
\eta \le 2 - d \ ,
\label{eq:detaeq}
\end{equation}
on the grounds that the density of states should diverge as the
transition is approached from the Bose glass side. In addition, since the
correlations must decay with distance at criticality, it follows from
Eq.~(\ref{eq:crscale}) that
\begin{equation}
d + z - 2 + \eta > 0 \ .
\end{equation}
Note that since $z > 0$ one can have a negative $\eta$ even in two
dimensions. There is also a general inequality applicable to random
systems~\cite{chayes86,chayes89}
\begin{equation}
\nu \ge {2 \over d} \ ,
\label{eq:exharris}
\end{equation}
which is a generalization of the Harris criterion~\cite{harris}.
The value of the dimension that should
be inserted into this expression is the number
of dimensions in which the system is random, i.e.\ the space dimension
$d$ and not $d+1$ or $d+z$.

As noted in the discussion
below Eq.~(\ref{aspect}) we need to know the dynamical
exponent $z$ in order to choose sample shapes which allow
a simple finite-size scaling analysis, i.e.\ the samples should
be of size $L \times L \times c L^z$, where $c$ is the aspect ratio.
Most of the simulations
were done assuming $z = 2$, the value predicted by Fisher et
al.~\cite{fisher89c}. We have done additional simulations with shapes
corresponding to other values of $z$, but find that the scaling is much
less good if $z$ is significantly different from 2. 
For $z=2$ we have taken 
two different aspect ratios 1/2 and 1/4, with the following systems
sizes: $4\times4\times8,\ 6\times6\times18$, and
$8\times8\times32$ for aspect ratio 1/2, and $6\times6\times9,\ 
8\times8\times16$, and $10\times10\times25$ for aspect ratio 1/4. 
We were unable to study larger lattices because the relaxation times
were too long.

As a test of our program we checked that we were able to reproduce the
results of Ref.~\onlinecite{cha91} in the absence of disorder and with
$\tilde\mu=0$. We found complete agreement~\cite{thesis} between the two
simulations.

\subsection{Equilibration}

\begin{figure}[htb]
\centering
\epsfysize=16 cm
\leavevmode
\epsffile{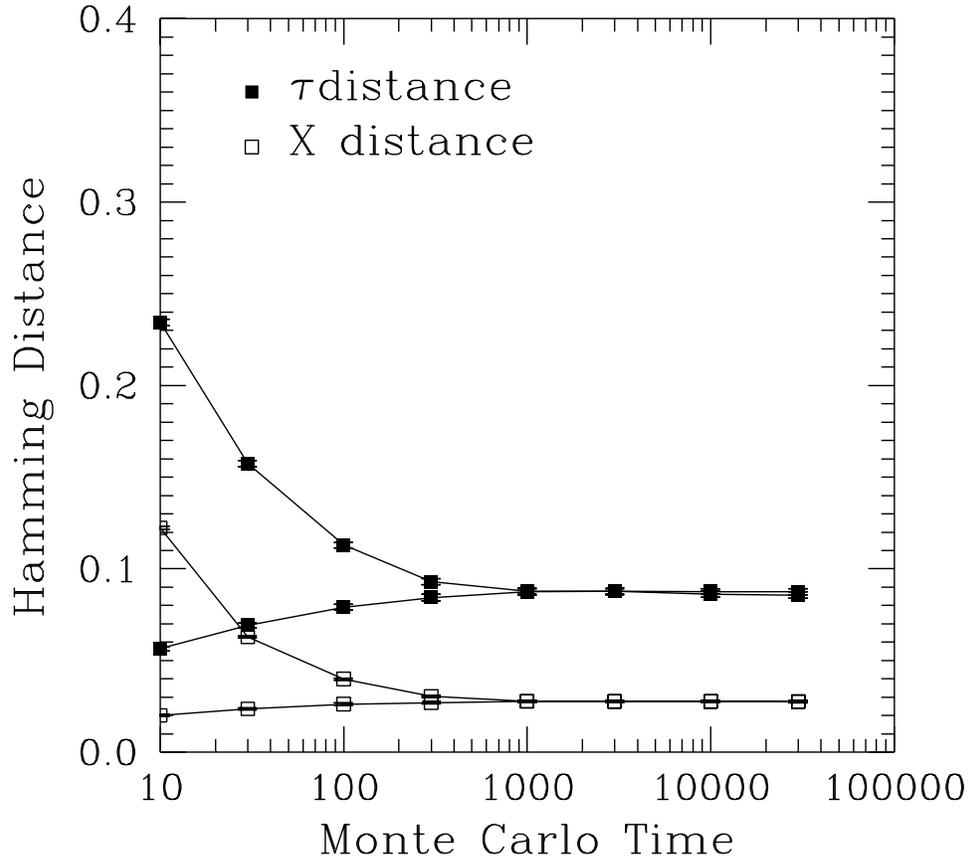}
\caption{
The $x$ and $\tau$ Hamming distances for a system of size 
$8 \times 8 \times 16$, with short-range interactions at the critical
coupling $K = 0.248$.
For each pair of curves, the upper one is for
$h_{\alpha,\beta}$ and the lower one for $h_\alpha$.}
\label{fig:deqquartxz}
\end{figure}

We test for equilibration using the method described in Section
\ref{sec:methods}. 
As an example,
Fig.~\ref{fig:deqquartxz}  shows the Hamming distance for the $x$ and
$\tau$ link variables for a system of size $8\times8\times16$,
at the critical point. We see that the system
equilibrates rather quickly in about 1,000 MCS. Also, we see that the
$x$ and $\tau$ link variables equilibrate in roughly the same time, as one
would expect since they are coupled through local moves.

\subsection{Determination of the Critical Point}

We start the analysis by locating the critical point.
Since, as discussed above,
we assume that $z=d=2$, the relevant quantity to plot, according to the
finite-size scaling analysis in subsection \ref{subsec:stiff},
is $\rho(0)L^2$.
Results for aspect ratio 1/4 
are shown in Fig.~\ref{fig:dquartcross}. 
Since the critical point
is located where the curves cross,
the figure demonstrates clearly that there is a transition close
to $K=0.25$ between 
a superfluid phase for $K > K_c$ with finite superfluid density,
$\rho_s$ (remember that $\rho_s\sim\rho(0)$), 
and an insulating phaseafor $K < K_c$ with zero superfluid density.
Our best estimate of the critical coupling is $K_c=0.248\pm0.002$. 
A substantial amount of computation went into the production of this
figure. Close to the critical point 1000 to 2000 disorder realizations were
performed, with, for the largest size,
an equilibration time of $T_0 = 10,000$ followed by 10,000
MCS for averaging with a measurement every 10 MCS. 

Simulations with aspect ratio $1/2$ were also performed and the
same critical coupling
was found, as expected since this is a bulk property.

\subsection{The Compressibility}
\begin{figure}[htb]
\centering
\epsfysize=16 cm
\leavevmode
\epsffile{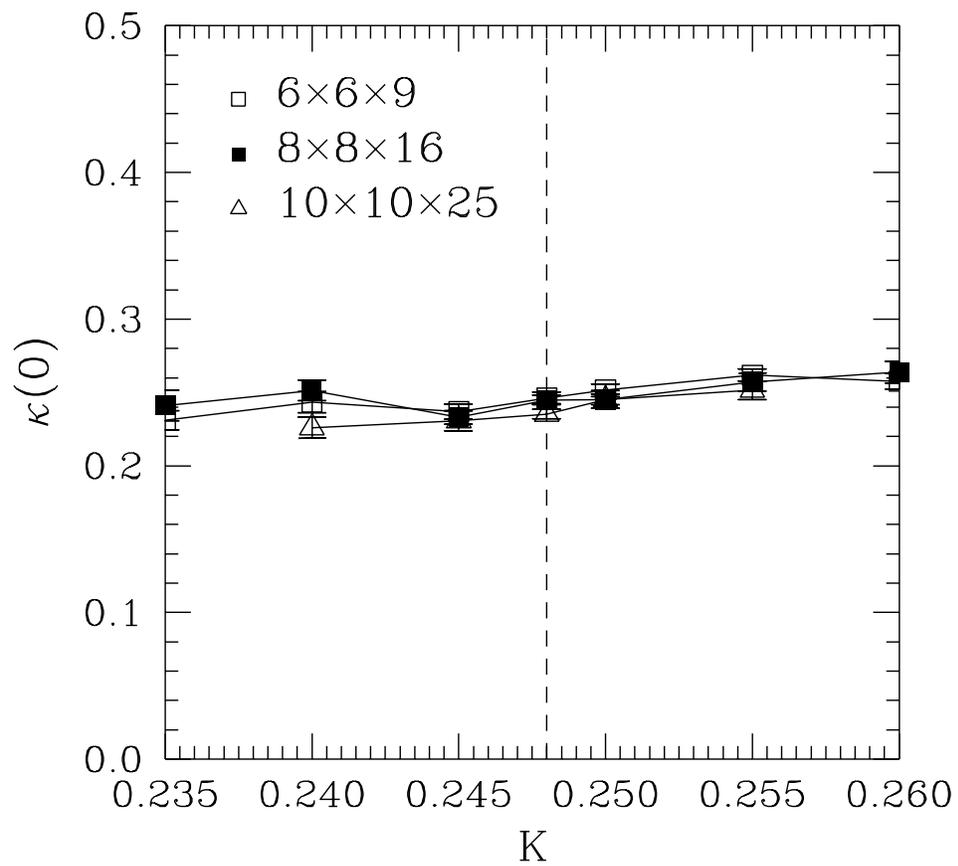}
\caption{
The compressibility at zero wave vector, $\kappa(0)$, for different
system sizes, as a function of
$K$ for short-range interactions. The critical point is at $K_c=0.248$.}
\label{fig:dkphalf}
\end{figure}

We now turn to the compressibility.  Fig.~\ref{fig:dkphalf}
shows the compressibility, as calculated from Eq.~(\ref{eq:kappa}),
for a range of different couplings
centered around the critical coupling $K_c=0.248$, for lattices with
aspect ratio 1/4.
We see that the compressibility remains finite through the
transition, including in the insulating phase, $K < K_c$.
This is consistent with the prediction that the insulating phase should be
a Bose glass with finite compressibility
in the presence of disorder~\cite{fisher89c}.
According to the scaling theory,
Eq.~(\ref{kappa:exponent}) a finite compressibility at criticality
implies $z = d\ ( = 2)$, as argued by Fisher et al.~\cite{fisher89c}.
By contrast, simulations
performed~\cite{thesis} with no disorder
and $\tilde\mu=0$, where the model becomes equivalent to a (2+1)D XY model,
find that the compressibility vanishes in the insulating phase,
consistent with it being a Mott insulator.

\begin{figure}[htb]
\centering
\epsfysize=16 cm
\leavevmode
\epsffile{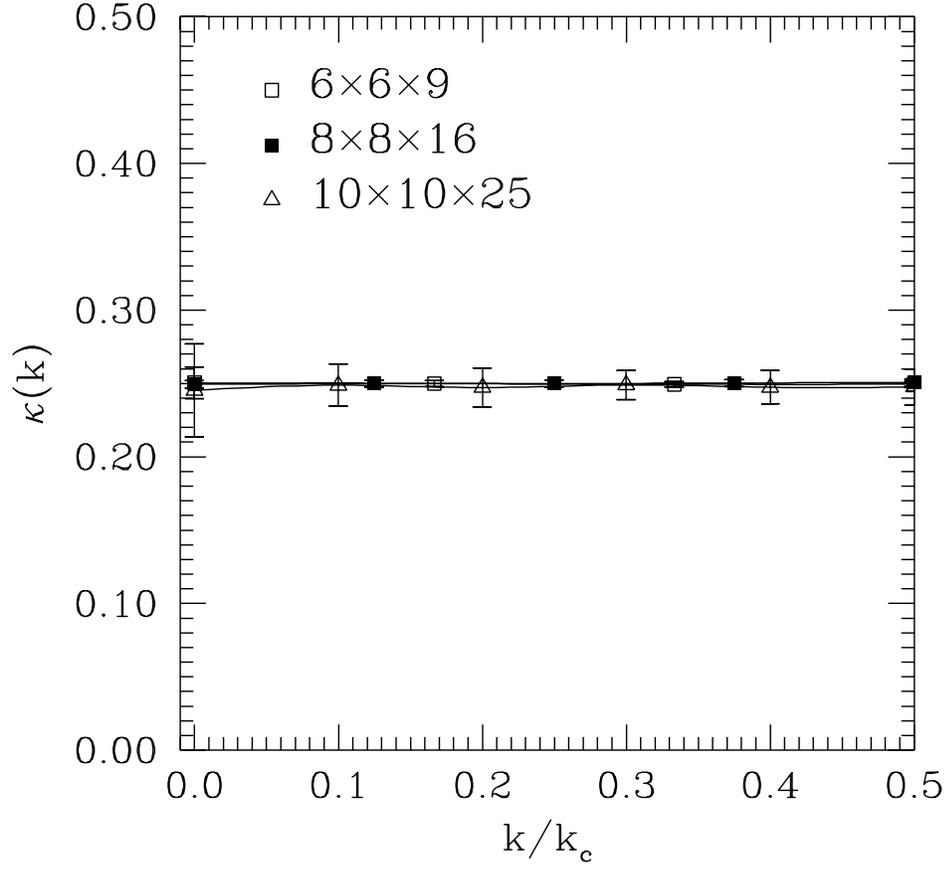}
\caption{
The compressibility of the model with short-range interactions
as a function of wave vector,
at the critical point, $K_c=0.248$,
for different system sizes. 
The solid lines are spline fits to the data points,
and $k_c=2\pi$.}
\label{fig:dkwquart}
\end{figure}

Fig.~\ref{fig:dkwquart}
shows the wave-vector dependent compressibility for the
aspect ratio
1/4, at the critical point $K_c=0.248$.
Similar results have been obtained for the aspect ratio 1/2.
Clearly there is no dependence on the
wave vector as expected from Eq.~(\ref{eq:kk}) and the result $z=d$.

From the above we have established that the insulating
phase is indeed a Bose glass and not a Mott insulator, at least
for the strength of the disorder that we have been considering
here, $\tilde\Delta=1/2$.
This is in agreement with previous
studies~\cite{krauth91b,singh,gold,runge}.

\begin{figure}[htb]
\centering
\epsfysize=16 cm
\leavevmode
\epsffile{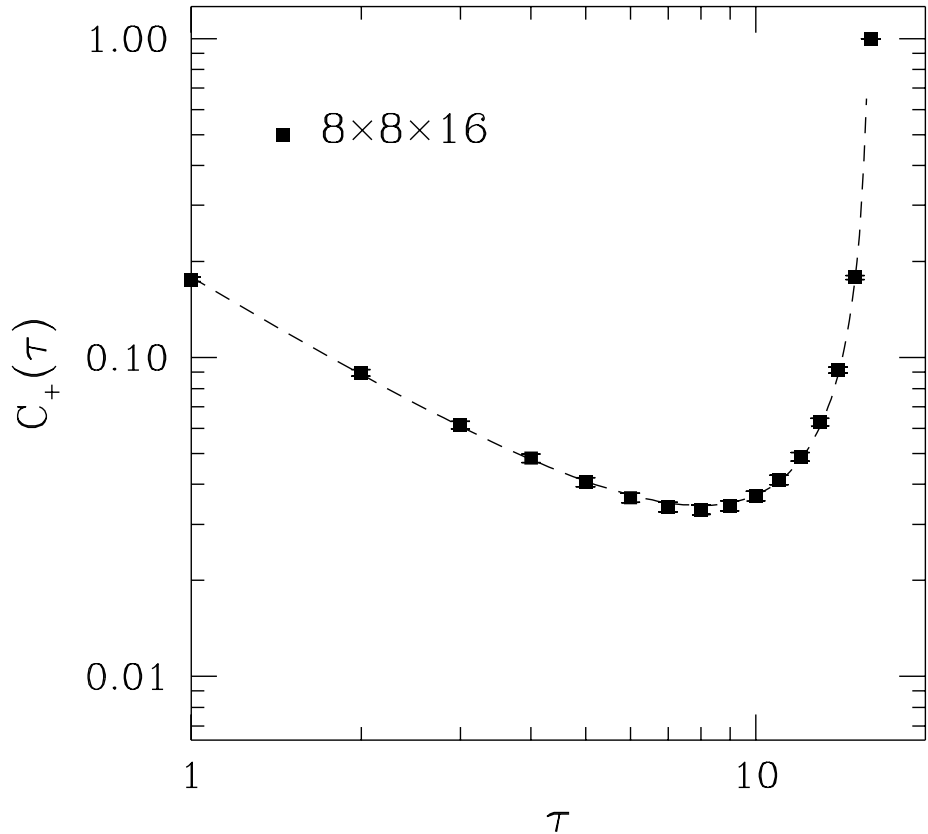}
\caption{
The correlation function (imaginary time Green's function) for positive
(imaginary) times, corresponding
to a boson propagating forward in time, 
for a system of size $8\times8\times16$ with short-range interactions, 
at $K=0.175$ which is far into the Bose glass phase.
The dashed line indicates a fit to the data of the form
$0.170(2)(\tau^{-1.10(8)}+(L_\tau-\tau)^{-1.10(8)})$.
The data for
$\tau > L_\tau / 2$ is actually the value of $C_-(L_\tau / 2 - \tau)$ as
discussed in the text. Thus the Green's function decays with a power of
time (rather than exponentially) in the Bose glass phase, as
predicted in Ref.~\protect\onlinecite{fisher89c}. }
\label{fig:dcorkp}
\end{figure}

As further evidence of the existence of the Bose glass we now turn to
a discussion of the correlation functions in the insulating
phase. One important prediction of the scaling
theory~\cite{fisher89c} is that the
Green's function should vary with a power of imaginary time,
\begin{equation}
C({\bf r}=0,\tau)\sim\rho_1(0)/\tau \ ,
\end{equation}
rather than exponentially, as might have been expected.
Here $\rho_1(0)$ is the single-particle density of states at zero
energy.
In order to check this prediction we did simulations deep in the
insulating phase, $K < K_c$.
Fig.~\ref{fig:dcorkp}
shows the time-dependent correlation function,
$C_+(\tau) \equiv C({\bf r}=0,\tau)$, for a system of
size $8\times8\times16$ at a coupling equal to
$K=0.175$, well below $K_c \simeq 0.248$.
The right hand part of the figure is obtained by calculating
$C_-(\tau)$, and using the relation $C_+(\tau)=C_-(L_\tau-\tau)$
discussed in subsection \ref{subsec:cf}.
For each disorder
realization, 30,000 MCS were performed, followed by
another 30,000 MCS to do the thermal averaging, and finally we averaged
over 100 different disorder realizations. The relatively elaborate
thermal averaging was done in order to obtain small error bars
at large $\tau$.
The dashed line is a power-law fit to the form
$0.170(2)(\tau^{-1.10(8)}+(L_\tau-\tau)^{-1.10(8)})$ where the
numbers in parentheses indicates uncertainties on the last digit.
This fit used all data points shown and gave a goodness of fit
of 0.84 and $\chi^2=7.9$. Here we define the goodness of fit
to be $\Gamma((N-2)/2,\chi^2/2)$, where $N$ is the number of
data points
and $\Gamma$ is the incomplete
gamma function. No sign of an exponential dependence on $\tau$
was observed.
The errors indicated are statistical and do not include 
possible systematic
errors. A fit at $K=0.15$ yielded a similar value for
the exponent of $1.05\pm0.04$.
We conclude that the time-dependent correlation functions
clearly display power-law behavior in the Bose glass phase
and, furthermore, the associated exponent
is close to 1 as predicted by scaling theory~\cite{fisher89c}.

\subsection{The Conductivity}

In the thermodynamic limit, $L \to \infty$, at vanishingly small $T$
($L_\tau \to \infty$), and for $\omega_n \to 0$, the conductivity at the
critical point 
should tend to a finite, universal value,
$\sigma^\ast$, as discussed in subsection \ref{subsec:cond}.
In the simulation there will be various corrections to this.
First of all, one might
ask whether the order of limits $T \to 0$ and
$\omega_n \to 0$ affects the value of $\sigma^\ast$,
even in the thermodynamic limit.
For the case of no disorder and
integer filling, where the transition is to the Mott insulator, the
answer is certainly yes~\cite{cha91}. In this case the conductance is
finite if the $T \to 0$ limit is taken first whereas $\sigma^\ast = 
\infty$ if 
one first takes the zero frequency (d.c.)
limit because a persistent current can
flow in the absence of umklapp processes, which vanish as $T \to 0$.
However, in the presence of disorder,
the d.c.\ conductivity is finite as $T \to 0$ and
so we see no reason why the order of the limits, $T \to 0$ and $\omega_n
\to 0$ should play a role for transition to the Bose glass phase
discussed here. We shall see that allowing for a dependence
on $\omega_n / T \propto \omega_n L_\tau$ does give a slighly better fit for
the case of short-range interactions, but
the value of $\sigma^\ast$ is not changed significantly.
Given the rather limited range of sizes that we can study, we feel that
the results are
consistent with there being
{\em no} dependence on $\omega_n / T$. One might also be concerned
about finite-size
corrections to the conductivity, but our results are consistent with 
their being very small. Of course, the conductivity is frequency
dependent and so will differ from the universal value when $\omega_n$
becomes comparable with some other scale, such as the ultraviolet cutoff
set by the lattice spacing.

\begin{figure}[htb]
\centering
\epsfysize=16 cm
\leavevmode
\epsffile{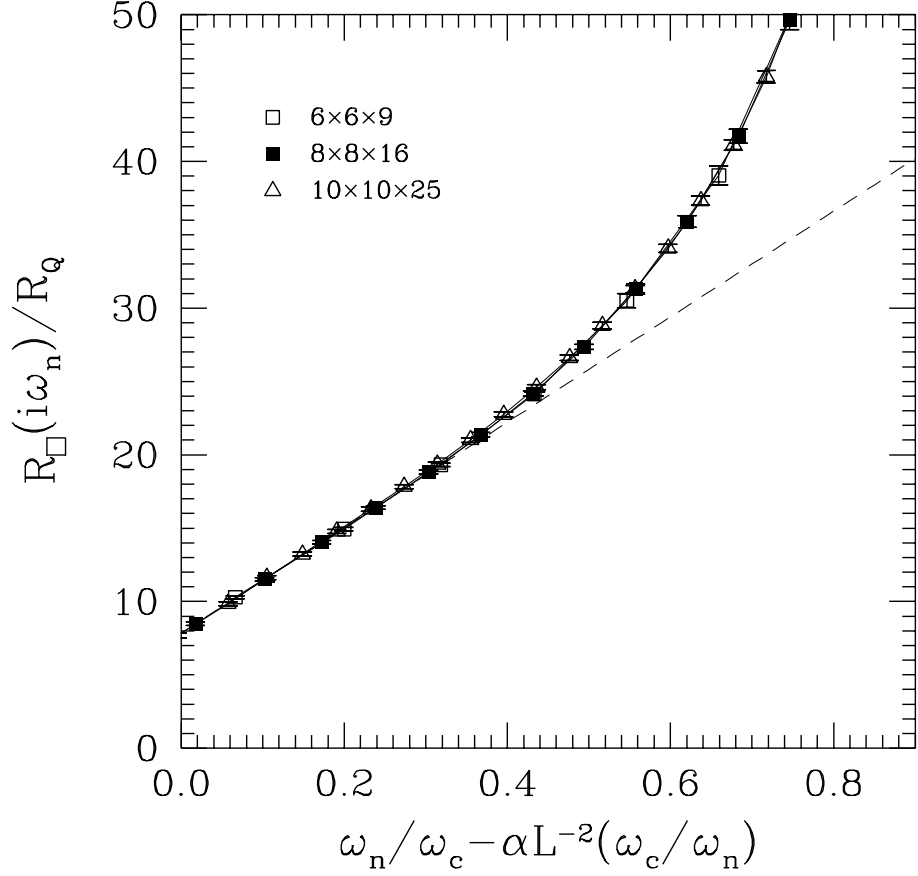}
\caption{
The resistivity in units of $R_Q=h/(2e)^2$, as a function of
$\omega_n/\omega_c-\alpha L^{-2}(\omega_c/\omega_n)$ for the model with
short-range interactions, where $\omega_c = 2 \pi$ and $\alpha = 0.179$.
The calculation was done at the critical
point, $K=0.248$. The aspect ratio was in this case 1/4, and the
system sizes shown were as indicated in the figure. The 
dashed line indicates a least square fit to the points with abscissa
less than 0.26 of the following form
$7.84(7)+34.9(5)(\omega_n/\omega_c-\alpha L^{-2}(\omega_c/\omega_n))$.
The correction involving the parameter $\alpha$ is proportional to
$T / \omega_n$ as discussed in the text. On physical grounds we expect
that $\alpha = 0$ in the thermodynamic limit, 
and indeed a fairly good fit, with almost the same
value of the d.c.\ resistivity, is obtained with $\alpha = 0$.}
\label{fig:dres}
\end{figure}

Fig.~\ref{fig:dres}  shows the resistance per square
(which is the same as the resistivity) plotted against
frequency~\cite{comment3}, evaluated
from Eq.~(\ref{eq:kubo})
for aspect ratio 1/4 at the critical point $K_c=0.248$. 
Again considerable computation has gone into the production of this graph
in order to obtain good statistics.
For the two smallest system sizes about 2,000
disorder configurations were generated while for the largest size
only 1,000 were done. 
From 3,000 to 10,000 MCS were done for equilibration
followed by the same number of sweeps for measurements.
The data collapse is excellent.

To determine the universal conductivity we have to analytically continue 
the MC
data to real frequency and extrapolate to $\omega=0$.
For typical quantum MC simulations, 
this analytic continuation is extremely difficult
to perform.
However it turns out to be straightforward in the present case, since
the data for the resistivity varies linearly at small $\omega_n$, which
implies,
\begin{equation}
\sigma(i\omega_n) = { \sigma^{\ast} \over (1+|\omega_n|\tau_0) } \ .
\end{equation}
This is easily seen to analytically continue to the Drude form of 
the conductivity:
\begin{equation}
\sigma(\omega+i\delta) = { \sigma^{\ast} \over (1-i\omega\tau_0) } \ .
\end{equation}
Thus the boson system at the critical point is neither insulator 
nor superfluid but rather a Drude metal.
The Drude parameter $\tau_0 \sim 1/\omega_c$
is a non-universal relaxation time controlled
in our model by the ultraviolet cutoff.

Assuming this linear variation of the resistivity
with $\omega_n$, a least squares fit
has a very small error.
The main source of error in the determination of the d.c.\ conductivity 
therefore comes from the uncertainty in the determination
of the critical point.
We estimated this error by making the same linear fit to the
resistivity data at the
ends of the interval given by the error bars of the critical coupling.
From all our data for two different aspect ratios we finally estimate
\begin{equation}
\sigma^\ast = (0.14 \pm 0.03) G_Q \ ,
\quad R_\Box^\ast = (7.4 \pm 1.6 ) R_Q \ .
\label{sigma:univ}
\end{equation}

The universal conductivity has previously been calculated for the case
of no disorder and integer boson filling~\cite{cha91},
where the insulating phase is a Mott insulator, rather than
the Bose glass discussed here.
In that case $\sigma^{\ast} \simeq 0.285 G_Q$. Thus we find
that even though the present model is somewhat more realistic,
including the disorder
takes us {\it further\/} from the experimental value which is in the
vicinity of unity.
A suitably defined universal conductance can be calculated
exactly in 1D~\cite{cha91} both with and without disorder.
The exact solution in 1D shows that the ratio of $\sigma^{\ast}$
in the dirty case and in the pure case is exactly 3/4.
This is of the same order of magnitude as
the ratio $0.14/0.285 \simeq 0.5$ between the MC results in 2D.
Hence we see that the trend of decreasing critical conductivity
upon adding disorder is the same as for the exact solution in 1D.

\subsection{The Exponent $\nu$}

\begin{figure}[htb]
\centering
\epsfysize=16 cm
\leavevmode
\epsffile{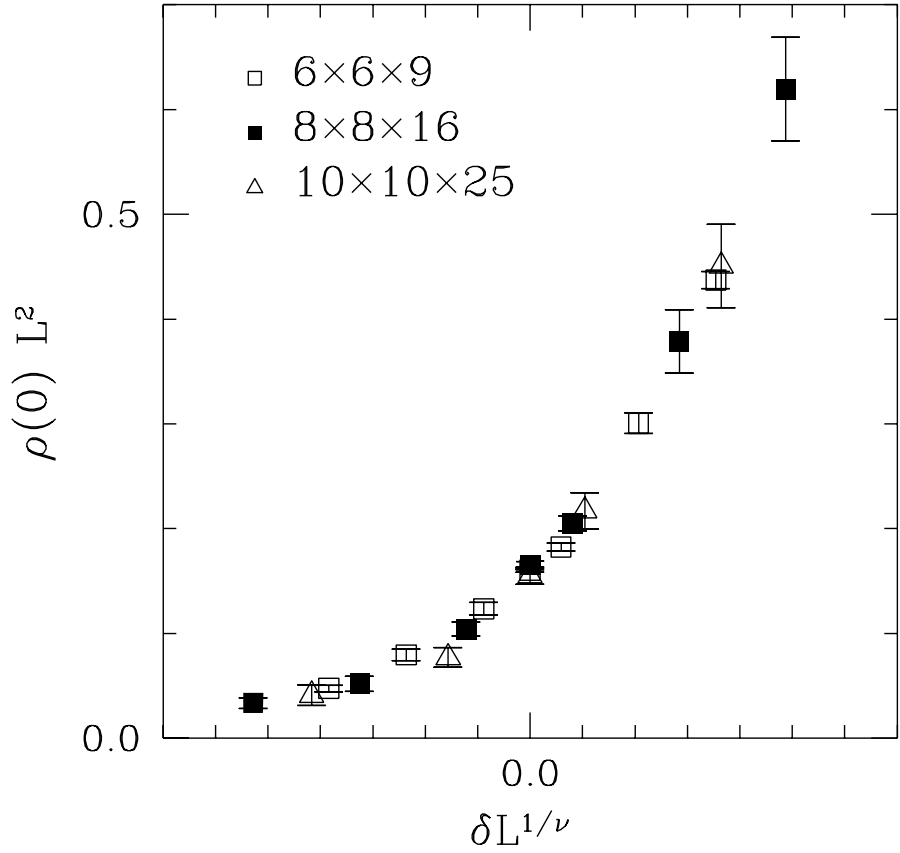}
\caption{
Scaling plot of $\rho(0)L^2$ versus $\delta L^{1/\nu}$, for short-range
interactions, where $\delta $
is the reduced coupling constant $(K-K_c)/K_c$ and $L$ is the
linear system size. The parameters used in
the plot are $K_c=0.248$, $\nu=0.90$.}
\label{fig:dscale}
\end{figure}

To determine the correlation length exponent we try to collapse the data
in a scaling plot of $\rho(0)L^2$ versus $\delta L^{1/\nu}$,
based on Eq.~(\ref{eq:rfinscale}).
The plot is shown in Fig.~\ref{fig:dscale},
for which the parameters used are $K_c=0.248$ and $\nu=0.9$.
From this and other plots for different values of the aspect ratio
we estimate 
\begin{equation}
\nu=0.90\pm0.10 \ .
\label{nu}
\end{equation}
Interestingly, the inequality, $\nu\ge2/d$, (with $d=2$ here) derived
by Chayes et al.~\cite{chayes86} is only just satisfied and may, in
fact, be an equality for this model. 
The equality, $\nu = 2/ d$, has been found for certain models of correlated
disorder~\cite{weinrib}.

\subsection{The Correlation Functions}

So far, we have determined the universal
values of $\nu$ and $\sigma^\ast$. We
have also found that the finite-size scaling works best with $z=2$ and,
according to the scaling theory,
our results for the compressibility agree with this.
In this subsection we conclude our discussion of the
model with short-range interactions by looking at the correlation
functions, which give us the value of the third exponent,
$\eta$, and another estimate for $z$.

\begin{figure}[htb]
\centering
\epsfysize=16 cm
\leavevmode
\epsffile{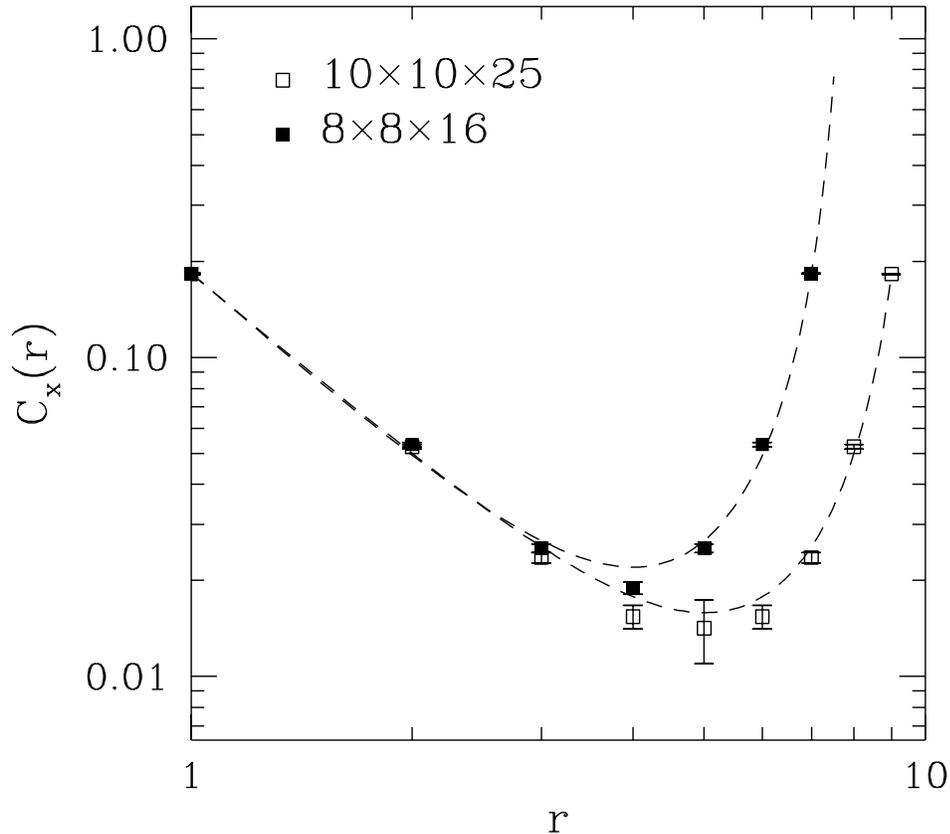}
\caption{
The equal-time correlation function of the $x$-link variables
as a function of spatial distance $r(=x)$ for short-range interactions.
The error bars
increase with increasing $r$ because the calculation involves
the average of the exponential
of a ``string'' of link variables, see Eq.~(\protect\ref{eq:cr}), which
can fluctuate hugely when the string is long. For this reason, the data
for $r > L/2$ were determined from $C_x(L - r)$.
Two data sets are shown,
for systems of size $L = 8$ and 10 with aspect ratio 1/4,
at $K=0.248$, which is the critical point.
The dashed line indicates fits to the data. For $L=8$ the fit is
$0.18(1)(r^{-2.02(1)}+(L-r)^{-2.02(1)})$.
For $L=10$ the fit is
$0.18(1)(r^{-1.94(2)}+(L-r)^{-1.94(2)})$.
}
\label{fig:dcxfive}
\end{figure}

The data for $C_x(r)$ for sizes $L=8$ and 10 with aspect ratio 1/4
at $K_c = 0.248$ 
are shown in Fig.~\ref{fig:dcxfive}.
In order to make full use of all the points we fit the
correlation functions to the following form
\begin{equation}
C_x(r) = c(r^{-y_x}+(L-r)^{-y_x}) \ ,
\label{yx}
\end{equation}
which takes the periodic boundary conditions into account. 
For $L=8$ the best fit had the
form $0.18(1)(r^{-2.02(1)}+(L-r)^{-2.02(1)})$.
For $L=10$ the fit was
$0.18(1)(r^{-1.94(2)}+(L-r)^{-1.94(2)})$.
These fits are indicated by the dashed lines in Fig.~\ref{fig:dcxfive}.

\begin{figure}[htb]
\centering
\epsfysize=16 cm
\leavevmode
\epsffile{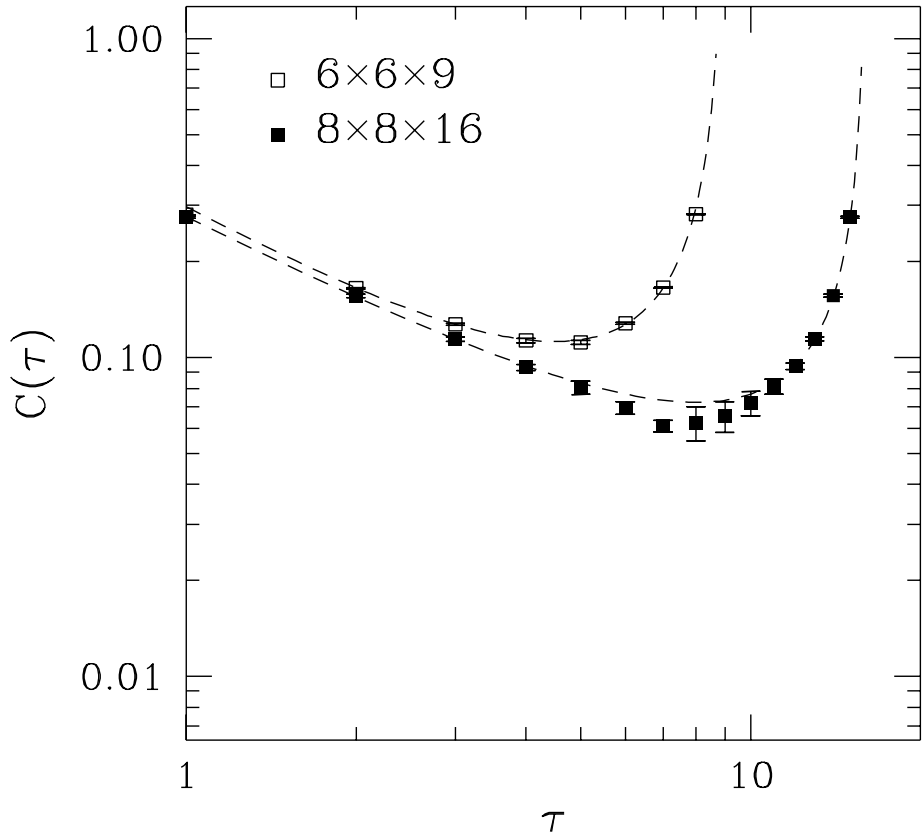}
\caption{
The correlation function (Green's function) for positive times
for short-range interactions for systems of size
$6\times6\times9$ and $8\times8\times16$, i.e.\ with aspect ratio 1/4.
The calculation was performed
at the critical point, $K_c=0.248$. 
The dashed line indicates fits to the data. For $L=6$ the fit is
$0.266(4)(\tau^{-1.03(1)}+(L-\tau)^{-1.03(1)})$.
For $L=8$ the fit is
$0.256(4)(\tau^{-0.94(1)}+(L-\tau)^{-0.94(1)})$.
The data for
$\tau > L_\tau / 2$ is actually the value of $C_-(L_\tau / 2 - \tau)$, as
discussed in the text.}
\label{fig:dczfour}
\end{figure}

In order to obtain $\eta$ and $z$ from Eq.~(\ref{eq:expo}) 
we also need to obtain results
for the time-dependent correlations. Since $\tilde\mu = 0.5$
there is statistical particle-hole symmetry~\cite{comment5} and so
$C_-(\tau)=C_+(\tau)$.
Fig.~\ref{fig:dczfour} shows the results at the critical
point $K_c=0.248$ for an aspect ratio of 1/4, where the data for
$\tau\ge L_\tau/2$ were obtained from $C_-(L_\tau-\tau)$.
The linear sizes were $L = 6$ and 8. We fit to the form 
\begin{equation}
c(\tau^{-y_\tau}+(L_\tau-\tau)^{-y_\tau}) \ ,
\end{equation}
and we find for $L=6$ the optimal fit has the
form $0.266(4)(\tau^{-1.03(1)}+(L-\tau)^{-1.03(1)})$.
For $L=8$ the best fit has the form
$0.256(4)(\tau^{-0.94(1)}+(L-\tau)^{-0.94(1)})$.
We see that the exponent governing the power-law
behavior is about 1/2 of the equivalent exponent in the space direction,
indicating from Eq.~(\ref{eq:expo}) that
the dynamical exponent $z$ must be close to 2. Combining all our estimates
for $z$ we find
\begin{equation}
z=2.0\pm0.1 \ .
\end{equation}

Our estimate for $\eta$ obtained from Eq.~(\ref{eq:expo}),
including results from the two aspect ratios, is
\begin{equation}
\eta=-0.1\pm0.15 \ .
\end{equation}
This agrees with the inequality 
Eq.~(\ref{eq:detaeq}) $\eta \le 2-d\ ( =0)$, 
which is possibly satisfied as an equality.
Noting that $\nu$ is given by Eq.~(\ref{nu}) and $\sigma^\ast$ by
Eq.~(\ref{sigma:univ}), this concludes our discussion of the universal
properties of the short-range model.

\section{Long-range coulomb interactions}\label{sec:lr}

We shall now include long-range Coulomb interactions.
Throughout this section we again take
\begin{equation}
\tilde\mu=\frac{1}{2},\;\;\tilde\Delta=\frac{1}{2} \ ,
\end{equation}
but, in addition, take the charge of the bosons to
be nonzero. Most of our studies used
\begin{equation}
{e^\ast}^2=\frac{1}{2} \ ,
\end{equation}
but we also have some results for ${e^\ast}^2=1/4$
as a check that the strength of the Coulomb term is irrelevant.
The long-range interactions force us to keep the total boson number
fixed, since it has to be compensated by a (fixed) background charge to
avoid an energy which is infinite in the thermodynamic limit.
We do not, therefore,
allow global moves in the time direction (the $\tau$-link variables).

As in the section above on short-range interactions we first discuss
what inequalities and estimates there are for the exponents. The result
$z = d$, quoted earlier is only applicable to short-range interactions.
For a $1/r$ potential, Fisher~\cite{fisher90b}, has argued that
\begin{equation}
z=1 \ .
\end{equation}
A simple way to see this~\cite{fisher:priv.comm}
is to compute the characteristic energy,
$\Delta\varepsilon$, 
given by the potential energy at $r=\xi$, 
i.e.\ $\Delta\varepsilon=G(r=\xi)\sim\xi^{-1}$,
where $G$ is the $1/r$ Coulomb potential.
If this is the relevant energy scale in the
problem then $\Delta\varepsilon\sim\xi^{-z}$ with z = 1.
This argument is trivially generalized to interactions falling off
with some arbitrary power of the distance, $G(r) \sim r^{-\lambda}$, and
leads to $z = \lambda$. We expect that this is valid for 
$\lambda$ smaller than $d$, the value for short-range 
interactions, and that for larger $\lambda$, the 
the dynamical exponent sticks at its short-range value, $z=d$.
Based on scaling of a renormalized charge 
Fisher et al.~\cite{fisher90a} derive the inequality
\begin{equation}
z\le 1 \ ,
\end{equation}
and an argument that the second sound velocity should not diverge at the
critical point~\cite{fisher89c} gives, quite generally,
\begin{equation}
z\ge 1 \ .
\end{equation}
Hence there is quite strong evidence that $z=1$ for the $1/r$
interaction. This is convenient for
the Monte Carlo work, because, although the computer time per update
increases by adding the long-range interaction, the number of lattice
points is not so large as for the short-range case
because we only have to scale $L_\tau$ with the
first power of $L$, rather than its square.

In the section above on short-range range interactions, the inequality
$\eta \le 2 - d$ was discussed. This was derived~\cite{fisher89c} with
the assumption that the density of single-particle states in the Bose
glass phase is finite at zero energy, an assumption
which is no longer correct with
$1/r$ interactions~\cite{shklovskii,efros}. In a classical model, 
called the ``Coulomb glass'', 
Efros and Shklovskii~\cite{shklovskii,efros} [ES] argue that that
the single-particle density of states,
$\rho_1(\varepsilon)$,
vanishes at $\varepsilon = 0$, due to the ``Coulomb gap''.
Assuming that 
\begin{equation}
\rho_1(\varepsilon) \sim \varepsilon^a \ ,
\label{cg:expt}
\end{equation}
ES obtain a bound on the density of states for small $\varepsilon$,
\begin{equation}
\rho_1(\varepsilon) \le  C \varepsilon^{d-1} \ , 
\end{equation}
so that
\begin{equation}
a \ge d - 1 \ .
\label{bgphase}
\end{equation}
The value of $a$ in 2D does not seem to be precisely known\cite{DLR}.
    
Since the Coulomb glass model is classical, the statistics of the
particles does not matter, so Eq.~(\ref{bgphase}) should be applicable
to the Bose glass phase, provided quantum fluctuations are unimportant
in this region, as is argued for the electron case~\cite{shklovskii,efros}.
It is then straightforward to determine the long time behavior of the
Green's function in the Bose glass phase. For $\tau > 0$ we have
\begin{eqnarray}
C(\tau) & = & \int_0^\infty d\varepsilon
e^{-\varepsilon |\tau|}\rho_1(\varepsilon) \\ \nonumber
& \sim & \frac{1}{\tau^{1+a}} \ .
\end{eqnarray}
The same argument, together with Eq.~(\ref{eq:expo}),
indicates that at criticality
$\rho_1(\varepsilon) \sim \varepsilon^{(d - 2 + \eta)/z}$.
We now assume, following Fisher et al.~\cite{fisher89c},
that $\rho_1(\varepsilon)$ at small $\varepsilon$
grows as the critical point is approached, 
in order to match onto
the delta-function density of states in the superconducting state. In
other words,
the density of states exponent is smaller at the critical point than in
the Bose glass phase, i.e.
$(d -2+\eta)/z  \le a $, or
\begin{equation}
\eta \le 2 - d + a z \ ,
\label{eta:coulomb}
\end{equation}
which, as expected, reduces to Eq.~(\ref{eq:detaeq}) for a constant density of
states, $a = 0$.
The bound $\nu\le2/d$, Eq.~(\ref{eq:exharris}), should also be valid
in the case of long-range interactions.

We now discuss the results from the simulations. Since there are strong
arguments, discussed above, that $z=1$, we work
with systems with shape
$L \times L \times L$ with $L = 6, 8, 10$ and 12.
In most cases we perform 3,000 MCS for equilibration followed by
3,000 MCS with a measurement every 10 MCS. Close to the critical point
the number of sweeps was generally larger for the larger sizes.
The number of disorder
realizations varied from 200 to 1000.
As was the case for short-range interactions we carefully check
for equilibration by computing the ``Hamming distances'' discussed in
Section \ref{sec:methods}.

\subsection{Determination of the Critical Point}
\begin{figure}[htb]
\centering
\epsfysize=16 cm
\leavevmode
\epsffile{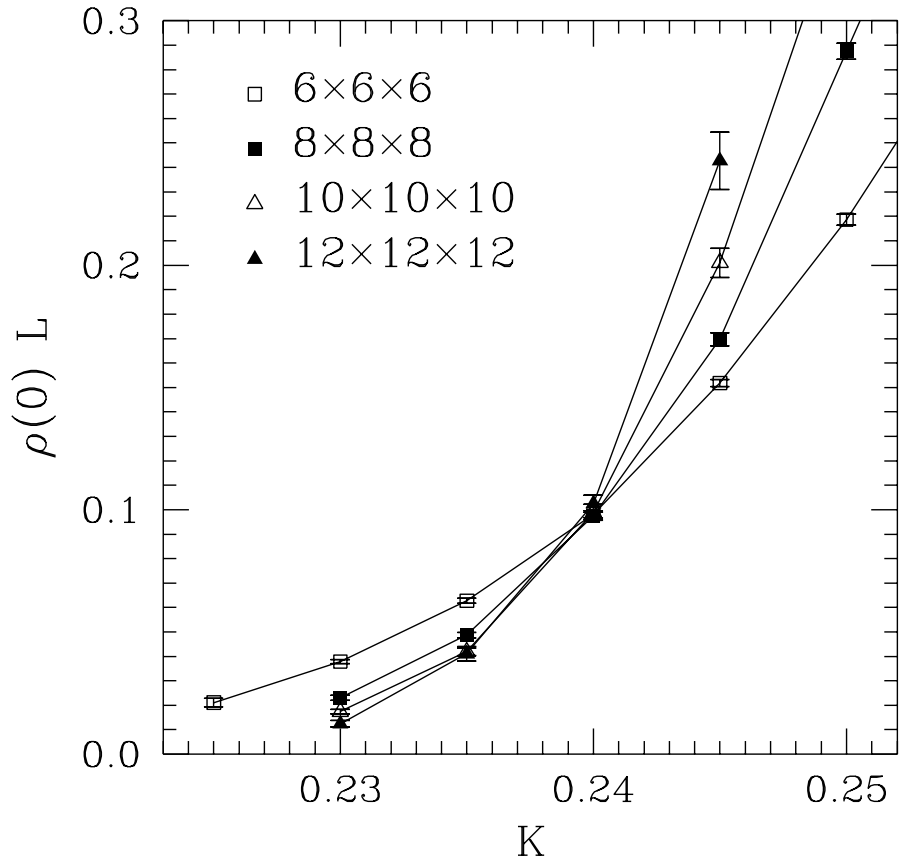}
\caption{
$\rho(0)L$ for the system sizes indicated, as a function of
$K$. The system sizes indicated correspond to an aspect ratio
of 1. The critical point, determined from the intersections,
is $K_c=0.240\pm0.003$.
The Ewald-sum form of the Coulomb potential was used with
${e^\ast}^2=1/2$.}
\label{fig:ccross}
\end{figure}

Since $z=1$ it follows from the discussion after Eq.~(\ref{aspect})
that we should look for the
intersections of data for $\rho(0)L$ against $K$ for different lattice sizes.
Our results are presented in Fig.~\ref{fig:ccross}
for the case of the Ewald-sum form of the potential with ${e^\ast}^2=1/2$.
Clearly the lines cross close to 
$K=0.240$ and, more precisely, we estimate the critical coupling to
be $K_c=0.240\pm0.003$, quite close to the value for the short-range
case. Since all four sizes intersect very close to the same point,
Fig.~\ref{fig:ccross} provides strong evidence that $z=1$, in agreement
with the scaling arguments.

An equivalent analysis can be performed with the Green's function form
for the potential, Eq.~(\ref{eq:green}),
with ${e^\ast}^2=1/4$.
The two forms of the potential are
different on short length scales,  and the 
values of ${e^\ast}^2$ are therefore not directly comparable. Although
we don't have enough
data for the largest size, $12\times 12\times 12$,
to perform a conclusive analysis, 
we can determine the critical coupling to
be $K_c \simeq 0.275$,
somewhat higher than for the Ewald form,
with reasonable certainty.

\subsection{The Conductivity}

\begin{figure}[htb]
\centering
\epsfysize=16 cm
\leavevmode
\epsffile{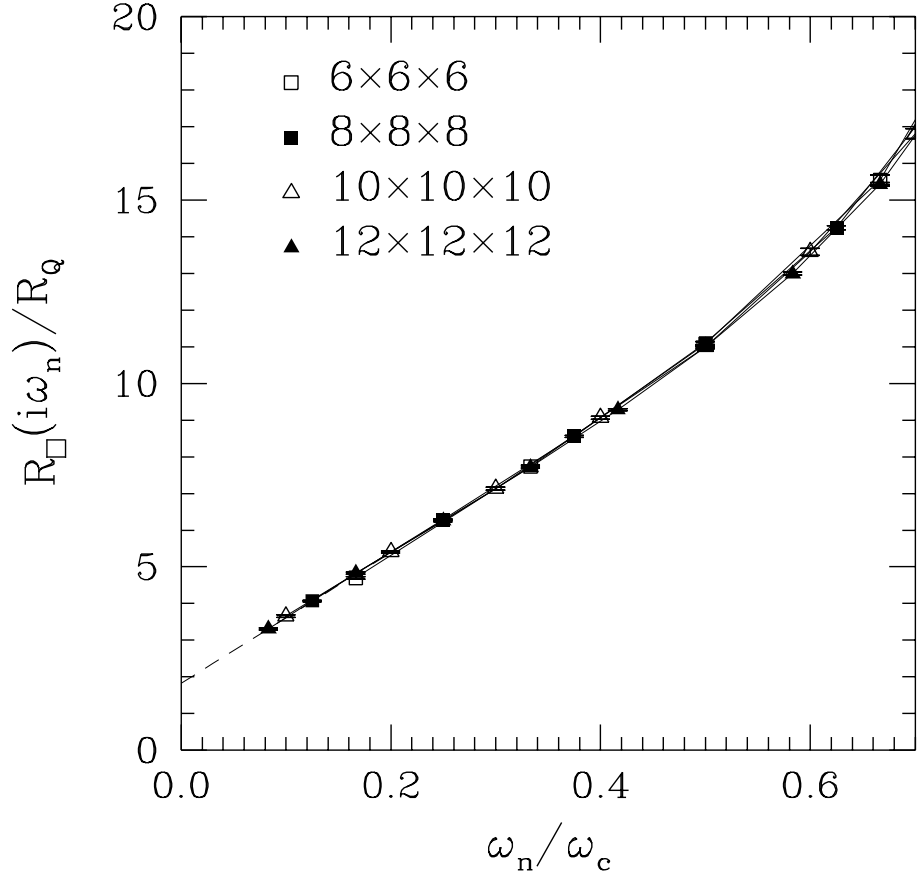}
\caption{
The resistivity in units of $R_Q=h/(2e)^2$, as a function of
$\omega_n/\omega_c$, where $\omega_c = 2 \pi$.
The calculation was done at the critical
point, $K=0.240$. The aspect ratio was in this case 1, and the
system sizes shown were as indicated in the figure. ${e^\ast}^2=1/2$ was
used along with the Ewald form for the potential. The
dashed line indicates a least square fit to the points with abscissa
less than 0.44 of the form
$1.82(2)+17.84(7)\omega_n/\omega_c$.}
\label{fig:cres}
\end{figure}

Following the approach used above for the short-range case, we plot, in
Fig.~\ref{fig:cres}, the resistivity, $R_\Box^\ast$
in units of $R_Q$, against frequency
at the critical point,  $K_c=0.240$. This data is for ${e^\ast}^2=1/2$,
with the Ewald method used to evaluate the Coulomb potential.
The collapse of the data is excellent, without any correction involving
$T/\omega_n$, which was used for the short-range
case.
This implies that we can interchange the
two limits $\omega\rightarrow 0$, and $T\rightarrow 0$ as expected. 
As for the short-range case, the data varies linearly with
$\omega_n$, implying a Drude form for the conductivity.
Making a least square fit to the data with
$\omega_n/\omega_c < 0.44$ we find
$R_\Box^\ast/R_Q$= 1.82(2).

\begin{figure}[htb]
\centering
\epsfysize=16 cm
\leavevmode
\epsffile{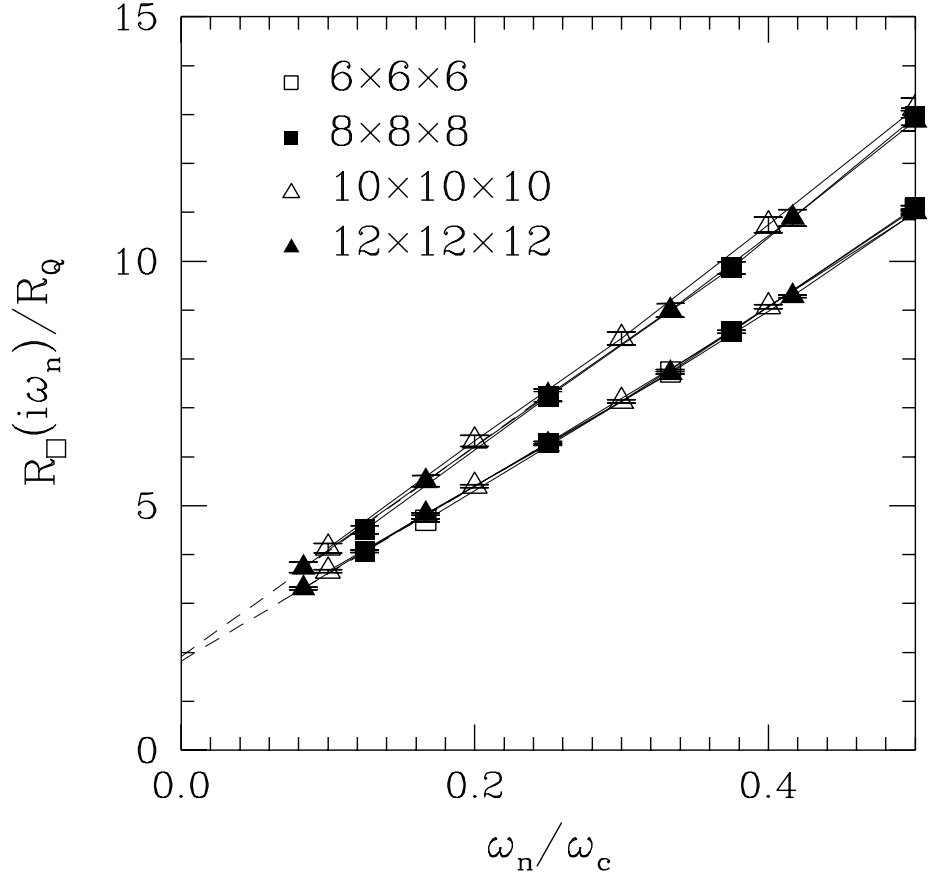}
\caption{The resistivity in units of $R_Q=h/(2e)^2$, as a function of
$\omega_n/\omega_c$, where $\omega_c = 2 \pi$.
For the upper curve the Green's-function form of the potential was used
with ${e^\ast}^2=1/4$, and an aspect ratio of 1.
The calculation was done at the critical
point for this potential, $K=0.275$.
The dashed line indicates a least square fit to the points with abscissa
less than 0.44 of the form
$1.91(7)+21.5(3)\omega_n/\omega_c$.
The lower curve is the results from the Ewald form of the potential
from Fig.~\protect\ref{fig:cres}. Although the results for the two forms
of the potential differ at finite frequency, they appear to extrapolate
to the same value in the d.c.\ limit, as expected since the d.c.\
resistivity is predicted to be universal.  }
\label{fig:cgreenres}
\end{figure}

We can now try to investigate the universality of $R_\Box^\ast$
by studying the model with ${e^\ast}^2=1/4$ and
the potential evaluated by the Green's function method. 
Fig.~\ref{fig:cgreenres} shows the resistivity at the critical point
$K_c=0.275$, along with the data already presented in
Fig.~\ref{fig:cres}.
We see two interesting things. Firstly, the
actual form of the resistivity
as a function of frequency is clearly different in the two cases.
However, the extrapolation to zero frequency is the same within the
uncertainties. For
the Green's function potential the best fit is
$R_\Box^\ast/R_Q = 1.91(7)$,
again for points
with abscissa less than 0.44. The agreement between
the zero-frequency limit of the two
sets of data in Fig.~\ref{fig:cgreenres} provides strong support for the
resistivity being universal at the transition.

We have also studied the effect of the aspect ratio on the
resistivity. This is important because
the aspect ratio is related to quantities
relevant for experiments, as we shall now show. An important concept in
mesoscopic physics is the phase coherence length
$\xi_{\rm inc}$. This length is expected to diverge as $T \to 0$, and
so, at criticality, should
be proportional to the Bose glass correlation length, $\xi$, making the
usual assumption that there is only one divergent length scale. To
determine how $\xi$ varies as a function of $T \sim L_\tau^{-1}$ note
that from finite-size scaling, the finite-size
relaxation time $\xi_\tau$ at criticality should
be proportional to $L_\tau$ and so characteristic lengths should scale as
$\xi_\tau^{1/z}$. Consequently $\xi_{\rm inc} \sim T^{-1/z}$, which
means that the the aspect ratio can be expressed as 
\begin{equation}
c   \equiv {L_\tau \over L^z}  \  \sim  {1 \over T L^z}
 \ \sim   \left({\xi_{\rm inc} \over L} \right)^z \ ,
\end{equation}
i.e.\ it is proportional to a power of the ratio of the phase coherence
length to the lattice size. Experiments are generally carried out in the
range
$L \gg \xi_{\rm inc}$ so one should view the conductivity as arising
from {\em incoherent} self-averaging of domains whose size is 
$\xi_{\rm inc}$. In the opposite limit, $L \ll \xi_{\rm inc}$, we expect
large variations from sample to sample 
(``bosonic universal conductance fluctuations''), 
and the average conductance
will not necessarily be the same as that obtained in the regime
$L \gg \xi_{\rm inc}$. Thus, the conductivity at zero frequency but
finite temperature is given by a scaling function,
$\widetilde\sigma(1/TL^z)$, where the argument is proportional to the
the aspect ratio. The experimental situation
corresponds to the limit of zero aspect ratio, whereas, the
simulations are done for a finite value.

We have therefore
performed calculations for two other aspect ratios, 1/2 and 3/2
respectively. In both cases we used the Ewald form of the potential with
${e^\ast}^2=1/2$. 
In order to obtain scaling plots for aspect ratios different from 1
it is necessary to
include corrections of the form $1 / L^2$.
Including this correction term, our estimates for $R_\Box^\ast$ agree
with those for aspect ratio unity, within the errors. Thus any 
dependence of $R_\Box^\ast$ on aspect ratio seems to be quite small.

In conclusion, we estimate the universal conductivity
at the critical point from all our data to be:
\begin{equation}
R_\Box^\ast=(1.82\pm0.20)R_Q \ , \;\; \sigma^\ast=(0.55\pm0.06)G_Q \ .
\end{equation}
No dependence on the aspect ratio, 
the microscopic form of the potential, 
the strength of the Coulomb interaction,
or particle density was observed.

\subsection{The Wave-Vector Dependent Compressibility}

\begin{figure}[htb]
\centering
\epsfysize=16 cm
\leavevmode
\epsffile{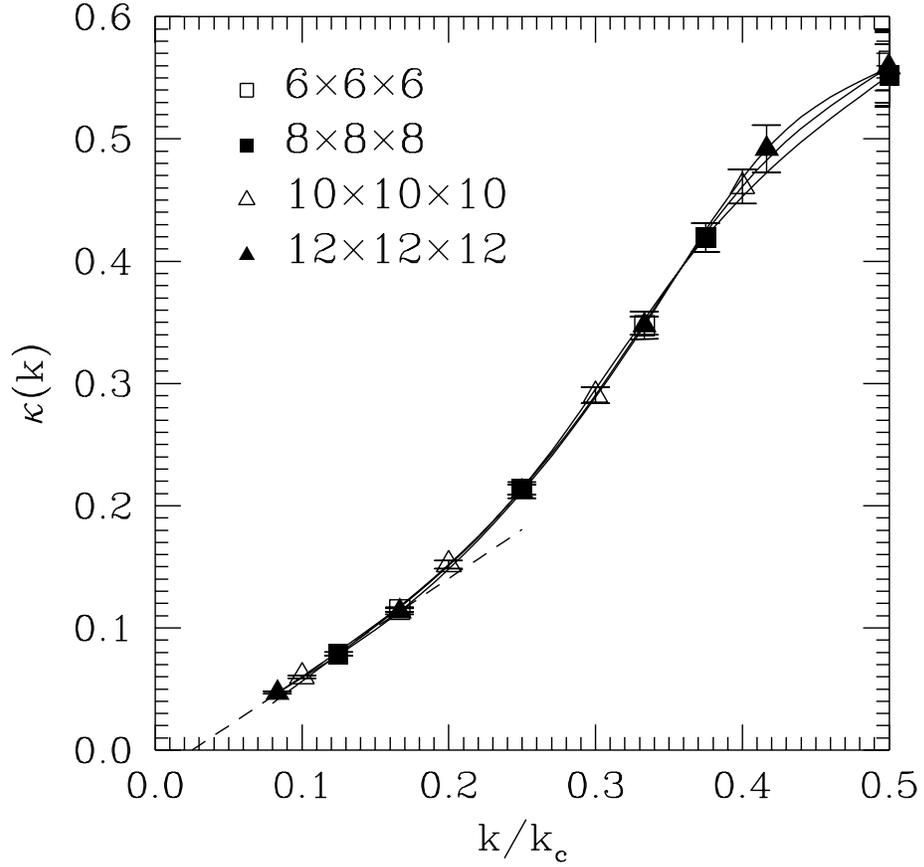}
\caption{
The compressibility at the critical point for the same
model as in Fig.~\protect\ref{fig:cres},
as a function of wave vector, for different system sizes, where
$k_c = 2\pi$ is the lattice cutoff. As expected
in Coulomb systems, the compressibility appears to
vanish as $k \to 0$.
The solid lines are spline fits to the data points, and $k_c=2\pi$.
Also indicated by a dashed line is a tentative fit to the
low frequency part with abscissa less than 0.19 of
the form
$-0.020(2)+0.8(2)k/k_c$.}
\label{fig:ckw}
\end{figure}

Because of the long-range interactions, the system is incompressible.
As a result the wave-vector dependent compressibility should vary
at criticality as
$\kappa(k) \sim k$ from Eq.~(\ref{eq:kk}) with $z=1$.
Fig.~\ref{fig:ckw} shows the data
at the critical point $K_c=0.240$, for aspect ratio 1, and
the Ewald form of the potential with ${e^\ast}^2=1/2$.
The solid lines shown are cubic splines
fitted to the data points.
The data for small $k$ seems to be roughly linear, as expected if $z=1$
but
one would need substantially smaller wave vectors to 
draw a firm conclusion.

Similar results results were obtained for aspect ratios $1/2$ and $3/2$.

\subsection{The Exponent $\nu$}
\begin{figure}[htb]
\centering
\epsfysize=16 cm
\leavevmode
\epsffile{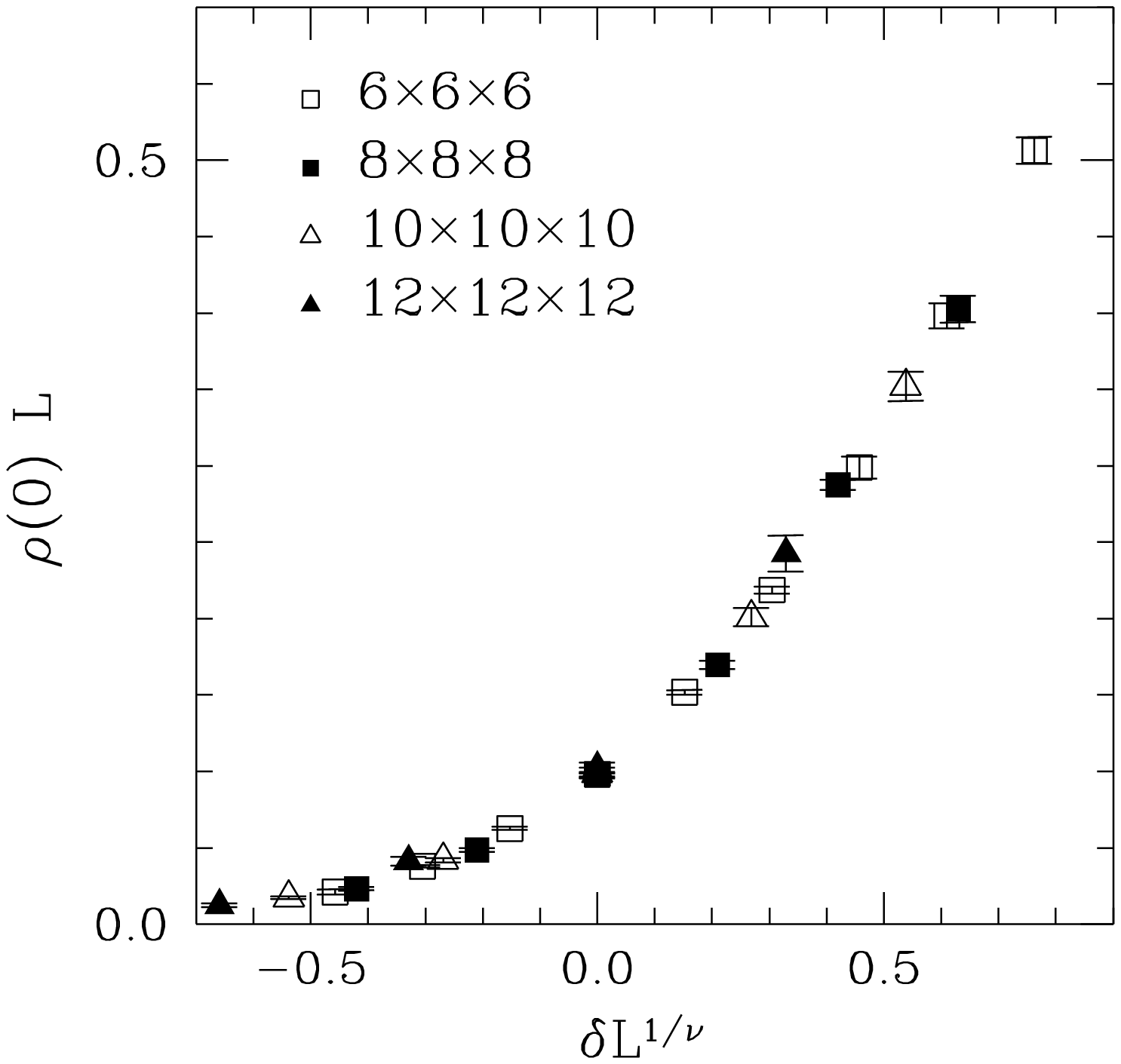}
\caption{
Scaling plot of $\rho(0)L$ versus $\delta L^{1/\nu}$, where $\delta $
is the reduced coupling constant $(K-K_c)/K_c$ and $L$ is the
linear system size. The model is the same as in
Fig.~\protect\ref{fig:cres}.
The parameters used in
the plot are $K_c=0.240$, $\nu=0.90$.}
\label{fig:cscale}
\end{figure}

To determine the correlation length exponent we try to collapse the data
in a scaling plot of $\rho(0)L$ versus $\delta L^{1/\nu}$,
based on Eq.~(\ref{eq:rfinscale}).
Fig.~\ref{fig:cscale} shows the data
for $\nu=0.90$ and $K_c=0.240$, with
aspect ratio 1, and
the Ewald form of the potential with ${e^\ast}^2=1/2$.
By considering all our data we estimate
\begin{equation}
\nu=0.9\pm0.15 \ .
\end{equation}
The estimate of $\nu$ is again consistent with the inequality
of Chayes et al.~\cite{chayes86}, $\nu\ge 2/d$, being
satisfied as an equality.

\subsection{The Correlation Functions}

Assuming $z=1$ we expect
the spatial correlation functions to have an asymptotic form of $C_x(r)\sim
r^{-y_x}$ with $y_x = 1+\eta$, see Eq.~(\ref{eq:expo}).
We have calculated the equal-time correlation functions 
for three different system sizes, $L \times L \times L$ with $L = 8, 10$
and 12. The
Ewald form of the potential was used with ${e^\ast}^2=1/2$,
and the calculation was performed
at the critical point $K_c=0.240$.
Despite there being quite a lot of noise in the data at
large arguments, we can fit to the form in Eq.~(\ref{yx}) 
with the result $y_x = 1.8\pm0.4$. Assuming that $z=1$ this then tells us
that $\eta=0.8\pm0.4$ in agreement with the previously derived
inequality, Eq.~(\ref{eta:coulomb}) $\eta \le az$, with $a \ge 1$.

In principle, the time-dependent correlation functions can also be determined.
However, unlike the equal-time correlation functions,
these involve injecting an extra particle
and then destroying it at a later time. 
Long-range interactions complicate the simulation of this correlation 
function and we have not attempted it.
Hence we cannot independently confirm the value $z=1$ of the dynamical 
exponent found in the scaling of $\rho(0) L^z$,
but believe it to be accurate.
Similarly, we have not attempted to obtain the
time decay of the correlation functions deep in the Bose glass phase,
which could give us information on the Coulomb gap exponent, $a$ in
Eq.~(\ref{cg:expt}).

\section{Discussion}\label{discuss}

We have investigated universal properties of
the $T=0$ Bose glass to superfluid transition in two
dimensions both for particles with short-range interactions and for
particles with long-range $(1/r)$ Coulomb interactions. We used a
version of the path
integral approach which corresponds to including only phase fluctuations
of the condensate.

For the case of long-range Coulomb interactions we find: 
$\nu = 0.90\pm 0.15,\ \eta = 0.8\pm 0.4,\ z \simeq 1.0$, 
and $\sigma^\ast = (0.55 \pm 0.06)\ G_Q$
where $G_Q^{-1} \equiv R_Q
\equiv h/(2e)^2 \approx 6.45$ k$\Omega$. 
This model should be in the
right universality class to describe the superconductor-insulator
transition in disordered thin films.
Experimental
results~\cite{haviland,liu,wang91,wang92,lee90,GOLDMANscaling} do not show
much support for the universality of $\sigma^\ast$, the values of
$\sigma^\ast / G_Q$ varying from about 0.6 to 2. However, it is possible that
many of these experiments were not at sufficiently low temperature to
probe the critical regime. We are not aware of any other theoretical
calculations with which to compare our results.

For the case of short-range interactions we find 
$\nu = 0.9 \pm 0.1,\ \eta = -0.10 \pm 0.15$, and $z = 2.0 \pm 0.10$. 
The universal conductivity is in this case 
$\sigma^\ast = (0.14 \pm 0.03) G_Q$.
These results are in reasonable accord
with those of Runge~\cite{runge}, who studied a hard-core boson model on
small lattices of size up to $4
\times 4$ by diagonalization, and obtained
$\nu = 1.4 \pm 0.3$ and $z = 1.95 \pm 0.25$. He also found $\sigma^\ast
/ G_Q = 0.16 \pm 0.01$, in agreement with our estimate, though
the error bar may be rather optimistic, since different assumptions for
the finite-size corrections led to significantly different values. 

\begin{figure}[htb]
\centering
\epsfysize=16 cm
\leavevmode
\epsffile{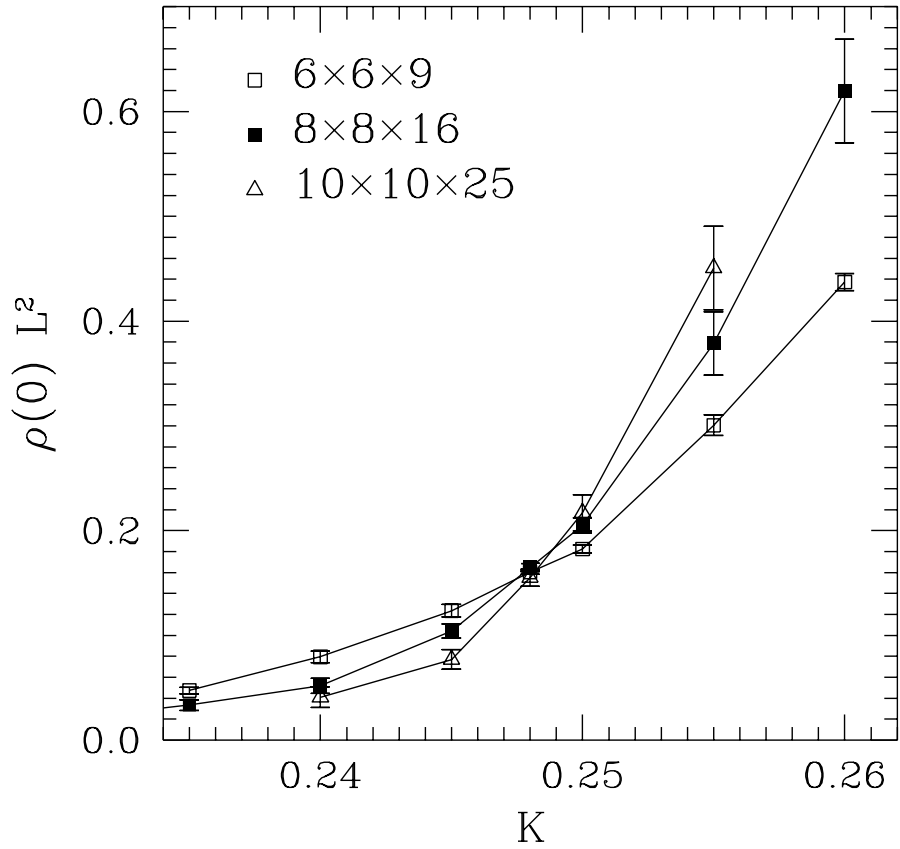}
\caption{
$\rho(0)L^2$ for the system sizes indicated, as a function of
$K$ for short-range interactions.
From the intersection of the curves we estimate
the critical coupling to be be $K_c=0.248\pm0.002$}
\label{fig:dquartcross}
\end{figure}

Since this work was completed, 
two groups have reported results for the short-range model which differ
from ours.
Batrouni et al.~\cite{batrouni93} have performed world-line
quantum Monte
Carlo simulations directly on the boson Hubbard model. They
find $\sigma^\ast=(0.45\pm0.07)\ G_Q$, but do not report values for the
exponents. We do not have an explanation for this discrepancy, though we
note that their data for $L^2\rho(0)$ do not splay out in the insulating
phase, $K < K_c$, as do ours (see Fig.~\ref{fig:dquartcross}), which makes
the location of the critical point harder.

Results which are totally
different from ours and also different
from those of Batrouni et al.~\cite{batrouni93} have
been found by Makivi\'c et al.~\cite{makivic93} who
performed world-line quantum Monte Carlo simulations on a hard-core Bose
system finding $\sigma^\ast=(1.2\pm0.2)\ G_Q$, with
$z=0.5\pm0.05$ and $\nu=2.2\pm0.2$. They used a large lattice of size
$64 \times 64 \times 48$ in the zero winding number sector,
and did finite-size scaling by considering sub-regions of sizes
$L_{\rm sub} \times L_{\rm sub} \times 48$ with $L_{\rm sub}$ between 
4 and 32.
Now the equilibration time varies as $L^{z_{\rm mc}}$, where, for this
model, the Monte Carlo dynamical exponent, $z_{\rm mc}$, 
is~\cite{MW-SMG} $\approx 6$.
It is therefore surprising to us that such a
large lattice can be equilibrated in the $3 \times 10^5$ sweeps that were
used. It is also unclear if their data would not have been consistent
with a larger dynamical exponent, had they scaled the 
the size of the sub-regions in the time direction like $L_{\rm sub}^z$
(and taken different values for $z$),
rather than leaving this size fixed at $L_\tau = 48$. Furthermore,
the two temperatures
used seem to be rather high since the critical coupling changed by a
factor of two between them.  Makivi\'c et al.~\cite{makivic93} propose
that the difference between their results and ours is that
amplitude fluctuations, neglected in our model but included in the boson
Hamiltonians, are relevant, so the two models are in
different universality classes. While this idea certainly can not be ruled
out, we don't yet feel that it has been conclusively demonstrated.
First of all, the only evidence for it is the numerical results, about
which we have some reservations discussed above. It
would be more compelling if there were additional evidence, such as a
calculation of the exponent for amplitude fluctuations, showing that
it is indeed relevant in the renormalization group sense. Furthermore, this
explanation does not explain the differences between the results of
Makivi\'c et al.~\cite{makivic93} and those of Batrouni et
al.~\cite{batrouni93} and Runge~\cite{runge}, which also included
amplitude fluctuations.

For the future, it would be very interesting to study the 
field-tuned transition~\cite{hebard90,hebard92} by Monte Carlo simulations,
since this is
expected to be in a different universality class~\cite{fisher90b}
from the disorder-tuned transition discussed here. The problem is that one
needs to find a representation of the problem in terms of a {\em real}
classical (D+1) dimensional 
effective Hamiltonian which incorporates both the magnetic field and
lack of microscopic particle-hole symmetry. Unfortunately, 
the phase representation, Eq.~(\ref{eq:isotropic}), though it
can be generalized to include a field and is still real for the
particle-hole symmetric case,
is complex in the absence of particle-hole symmetry, and the
link representation used here becomes complex in the presence of a
field, as can be seen from Eq.~(\ref{vectorpot}). We
are therefore unaware of any representation of the problem suitable for
Monte Carlo simulations.

\acknowledgements

We thank 
M.~P.~A.~Fisher,
A.~Hebard,
R.~Dynes,
J.~M.~Valles,
A.~M.~Goldman,
M.-C.~Cha,
K.~Runge,
N.~Trivedi,
G.~Zimanyi,
I.~Affleck,
J.~Gan,
M.~Gingras,
H.~J.~Schulz,
R.~Scalettar,
and S.~Kivelson
for a large number of valuable suggestions and discussions.
APY and ESS acknowledges the kind hospitality of Ecole Normale
Sup\'erieure, Paris, and CEA, Saclay, 
where a major part of this work was done,
as well as generous cpu-time allocations at the CCVR, Saclay, and
San Diego Supercomputer Cray facilities.
ESS also acknowledges support from the Danish Natural Research Council,
as well as from NSERC and CIAR of Canada.
MW is supported by the Swedish Natural Science Research Council.
SMG is supported by DOE grant DE-FG02-90ER45427 and 
NCSA Grant No.~DMR-910014N.
APY and ESS are supported by NSF DMR-91-11576.

\newpage

%

%
%

%
%

%
%
%
%
%
%
%

%
%

%
%

%
%


\begin{references}
\bibitem{comment-1}
The criterion for weak disorder being relevant is~\cite{harris}
$\nu < 2/d$, where $\nu$ is the correlation length of the pure system
and $d$ the dimensionality. At the Kosterlitz-Thouless
transition~\cite{KT}, $\nu
= \infty$, corresponding to an exponential, rather than power-law,
divergence, so disorder is highly irrelevant.
\bibitem{comment5}
Statistical particle-hole symmetry occurs when $\tilde\mu = n/2$ with
$n$ an integer. Microscopic particle-hole symmetry, which means that
Hamiltonian is invariant under $ J^\tau \to - J^\tau$ for 
a single sample, occurs when there is no disorder and $\tilde\mu$ is an
integer (or a half-integer), i.e.\ the boson filling is an
integer (or a half-integer). The Hamiltonian in the phase representation
is only real for no disorder and integer filling.
\bibitem{comment2}
It might be thought, from the definitions,
Eqs.~(\ref{cplustau}) and (\ref{cminustau}), 
that $C_+(\tau) = [C_-(\tau)]^\ast$. 
However, this is not correct. Both correlation functions
are real, but different from each other except when there is statistical
particle-hole symmetry. The point is that, except with integer filling
and no disorder (for which there is exact particle-hole symmetry),
the Hamiltonian is not real in the phase representation, 
so taking the complex conjugate is not trivial.
\bibitem{comment3}
The horizontal axis is really $\omega_n / \omega_c -
\alpha L^{-2}(\omega_c / \omega_n)$, where $\alpha = 0.179$
and $\omega_c = 2 \pi$ is the lattice cutoff. The correction
factor involving alpha is proportional to $T/\omega_n$ as discussed in
the text. The value of $\alpha$ is
considerably smaller than the best fit for the case of no
disorder~\cite{cha91} which gave $\alpha = 0.74$.
The data scale fairly well even for $\alpha = 0$ and this fit gives a
very similar value of $\sigma^\ast$.
As discussed in the text, it is expected that the correct value is
$\alpha = 0$.
\end{references}
\end{document}